\documentclass[journal=jacsat,manuscript=article]{achemso}
\usepackage[normalem]{ulem}
\usepackage[version=3]{mhchem} 
\usepackage[dvipsnames]{xcolor}

\author{Anubhab Sahoo}
\affiliation[Chennai]
{Department of Physics, Indian Institute of Technology Madras, Chennai -600036, India.}

\author{Tejendra Dixit}
\affiliation[IIITDM Kancheepuram, Chennai]
{Optoelectronics and Quantum Devices Group, Department of Electronics and Communication Engineering, Indian Institute of Information Technology Design and Manufacturing Kancheepuram, Chennai-600127, India.}

\author{K. V. Anil Kumar}
\affiliation[Chennai]
{Department of Physics, Indian Institute of Technology Madras, Chennai -600036, India.}

\author{K. Lakshmi Ganapathi}
\affiliation[2D IIT]
{2D Materials Research and Innovation Group, Indian Institute of Technology Madras, Chennai-600036, India}
\alsoaffiliation[QuCenDiEM]
{Quantum Center of Excellence for Diamond and Emergent Materials (QuCenDiEM) group, Indian Institute of Technology Madras, Chennai-600036, India}
\alsoaffiliation[NIT]
{Department of Physics, National Institute of Technology Kurukhetra, Kurukhetra-136119, India.}

\author{Pramoda K. Nayak}
\affiliation[Chennai]
{Department of Physics, Indian Institute of Technology Madras, Chennai -600036, India.}

\alsoaffiliation[2D IIT]
{2D Materials Research and Innovation Group, Indian Institute of Technology Madras, Chennai-600036, India}
\alsoaffiliation[jain]
{Centre for Nano and Material Sciences, Jain (Deemed-to-be University), Jain Global Campus, Kanakapura, Bangalore-562112, India}

\author{M. S. Ramachandra Rao}
\affiliation[Chennai]
{Department of Physics, Indian Institute of Technology Madras, Chennai -600036, India.}
\alsoaffiliation[QuCenDiEM]
{Quantum Center of Excellence for Diamond and Emergent Materials (QuCenDiEM) group, Indian Institute of Technology Madras, Chennai-600036, India}

\author{Sivarama Krishnan}
\email{srkrishnan@iitm.ac.in}
\phone{+91 (0)44 2257 4857}
\affiliation[Chennai]
{Department of Physics, Indian Institute of Technology Madras, Chennai -600036, India.}
\alsoaffiliation[QuCenDiEM]
{Quantum Center of Excellence for Diamond and Emergent Materials (QuCenDiEM) group, Indian Institute of Technology Madras, Chennai-600036, India}
\title[An \textsf{achemso} demo]
 {Elucidating the Role of Electron Transfer in the Photoluminescence of $\boldsymbol{\mathrm{MoS_{2}}}$ Quantum Dots Synthesized by fs-Pulse Ablation}

\abbreviations{fs-PLAL, $\mathrm{MoS_{2}}$ QDs, PL, UV, XPS, TRPL}

\begin{document}



\newpage
\begin{abstract}

Herein, $\mathrm{MoS_{2}}$ quantum dot (QDs) with controlled optical, structural, and electronic properties are synthesized using the femtosecond pulsed laser ablation in liquid (fs-PLAL) technique by varying pulse-width, ablation power, and ablation time to harness the potential for next-generation optoelectronics and quantum technology. Furthermore, this work elucidates key aspects of the mechanisms underlying the near-UV and blue emission, the accompanying large Stokes-shift, and the consequent change in sample color with laser exposure parameters pertaining to $\mathrm{MoS_{2}}$ QDs. Through spectroscopic analysis, including UV-visible absorption, photoluminescence, and Raman spectroscopy, we successfully unravelled the mechanisms for the change in optoelectronic properties of $\mathrm{MoS_{2}}$ QDs with laser parameters. We realize that the occurrence of a secondary phase, specifically $\mathrm{MoO_{3-x}}$, is responsible for the significant Stokes-shift and blue emission observed in this QDs system. The primary factor influencing these activities is the electron transfer observed between these two phases, as validated by excitation dependent photoluminescence, XPS and Raman spectroscopies.
\end{abstract}


\newpage
\section{Introduction}
The field of optoelectronics and nanophotonics has experienced significant intrigue with quantum dots (QDs) of semiconducting systems due to their ability to provide a versatile platform for modulating and establishing electronic and optical characteristics in an intuitive way \cite{Alivisatos933,Nirmal1999,Yoffe2002}.
This has been leveraged in a wide range of applications, from optoelectronics to biomedical \cite{Arakawa2002,JIN2015,Hildebrandt2017,MANIKANDAN2019,Yue2019,Cotta2020}, with a particular focus on quantum electronics, quantum light-emitting diodes (QLED), single-photon detectors and biosensors 
\cite{Arup2017,Bankar2017,Mukherjee2016,Guo2020}. Important requirements for these applications are QDs with high quantum yield (QY) \cite{Yin2014,Arup2017}, photostability \cite{Arup2017}, and tunable bandgap. These are all easily implementable in QDs based on transition metal dichalcogenides (TMDCs). The key to each of these realizations lies in the synthesis procedures that have the capability to generate QDs with tunable size \cite{Rounder2006}, shape, surface defects, and hybrid structures \cite{Mukherjee2016,Yue2019}. Owing to its ease of preparation, one of the most stable and emerging TMDC, $\mathrm{MoS_{2}}$ in zero-dimensional form, exhibits maximum confinement effect, resulting in extended absorption spectra ranging from visible to UV regime \cite{Splendiani2010}. Further, the high surface-to-volume ratio of $\mathrm{MoS _{2}}$ makes it more photosensitive than its monolayer counterpart, with PL emission QY up to the order of $10^{-1}$ ($22\ \%$) with enhanced light coupling \cite{NGUYEN2019,Asha_2023}. In addition, as a biocompatible QD, photostability makes it substitute for toxic inorganic compounds of groups II-VI (CdSe, ZnSe, CdTe, etc.) and long-range organometallic framework-based materials for photocatalytic and bio-imaging applications. \cite{GaoStokes_2017,Jaiswal_Dual_2022}\\

 $\mathrm{MoS_{2}}$ QDs have been explored to understand and study the optical properties, i.e. absorption, PL emission, and transient PL (TRPL), by scientific groups, specially in the UV regime. From prior studies, irrespective of synthesis methods, $\mathrm{MoS_{2}}$ nanoparticles below 10 nm show strong confinement effects for samples made via both bottom-up and top-down approaches \cite{Lin2015,Portone2018,Du_RSC2015,Wen_ASS_2015,Ali2022,LAMBORA2023}. Among other optical properties, high-efficient PL emission in the blue range brings attention as it features a high QY and large Stokes shift. QDs with large Stokes shift ($>$ 70 nm) play an important role in improving the signal-to-noise ratio in bio-imaging by avoiding cross-talk between PL excitation and emission \cite{GaoStokes_2017}. The origin of these photoluminescent emissions is linked to surface defect states present in $\mathrm{MoS_{2}}$, which include previously reported factors such as oxygen adsorbed groups and sulfur defects \cite{Gan_2015,Li_SR_2017,NGUYEN2019,Asha_2023}. In colloidal form, synthesis via chemical routes hinges on the selection of a suitable solvent, influencing and functionalizing the characteristics of surface groups \cite{Lin2015,Yin2019,Asha_2023}. Furthermore, the interface between the QD's surface and solvent significantly influences the emission center, as it is affected by the electron affinity and polarity of the dispersing medium \cite{Asha_2023}. Previous findings on fs-PLAL synthesized $\mathrm{MoS_{2}}$ QDs prioritized the synthesis technique and a single control parameter \cite{Gan_2015,Li_SR_2017,Xu_SR_2019,SUNITHA2018}. 
The major contribution for the PL emission in this method is reported as the surface functional groups, which also enhance the PL QY \cite{Xu_SR_2019,NGUYEN2019}. In most instances of PL emission, observed substantial Stokes shift, typically 80-100 nm away from the PL excitation wavelength, necessitates a thorough and adequate explanation. While various approaches are being employed to investigate the PL emission properties of $\mathrm{MoS_{2}}$, there is currently a lack of systematic and simultaneous exploration of the gradual development of surface states in a single comprehensive study. \\

Herein, we emphasize the following factors for investigating the PL emission in the range of 400-450 nm of these $\mathrm{MoS_{2}}$ QDs: (a) the relation of the PL emission with absorption edge, which is responsible for it, (b) cause of large Stokes shift in PL emission spectra, and (c) which surface state or phase it forms during the synthesis process. To gain insight into these phenomena, we synthesized $\mathrm{MoS_{2}}$ QDs with controllable surface states through a single-step, chemical-free approach using fs-PLAL in deionized water. By altering the ablation parameters, specifically the average power and ablation duration, while maintaining an optimized pulse width of 600 fs, we ablated $\mathrm{MoS_{2}}$ pellets to obtain QDs with intriguing optical properties.  A detailed explanation of the experimental setup and synthesis process is explained in Figure S1 (supporting information (SI)). The surface states are probed with Raman study and XPS, which reveal the formation of an $\mathrm{MoS_{2}}$/$\mathrm{MoO_{3-x}}$ phase. This is also evidenced by the UV-visible absorption study, which suggests the formation of a secondary phase in the UV band along with two additional NIR absorption peaks that turn the sample colour from pale yellow to blue. In this work, we unambiguously establish the origin of the blue emission in PL spectra to the formation of an additional $\mathrm{MoO_{3-x}}$ phase along with the $\mathrm{MoS_{2}}$ in these QDs, correlated to the parameters employed in this generic synthesis process. The existence of the $\mathrm{MoO_{3-x}}$ phase within these QDs results in an electron transfer from the excited state of the $\mathrm{MoS_{2}}$ phase to the $\mathrm{MoO_{3-x}}$ phase, ultimately giving rise to blue emission. We have substantiated this phenomenon through our investigation employing excitation-dependent PL and TRPL studies. To the best of our knowledge, this work is among the first to uncover the hitherto unforeseen role of the interplay between the two phases in this important TMDC system on the nanoscale.\\

\section{Result and discussion}
\begin{figure}[h!]
\renewcommand{\thefigure}{1}
 \centering
\includegraphics[width=0.9\linewidth]{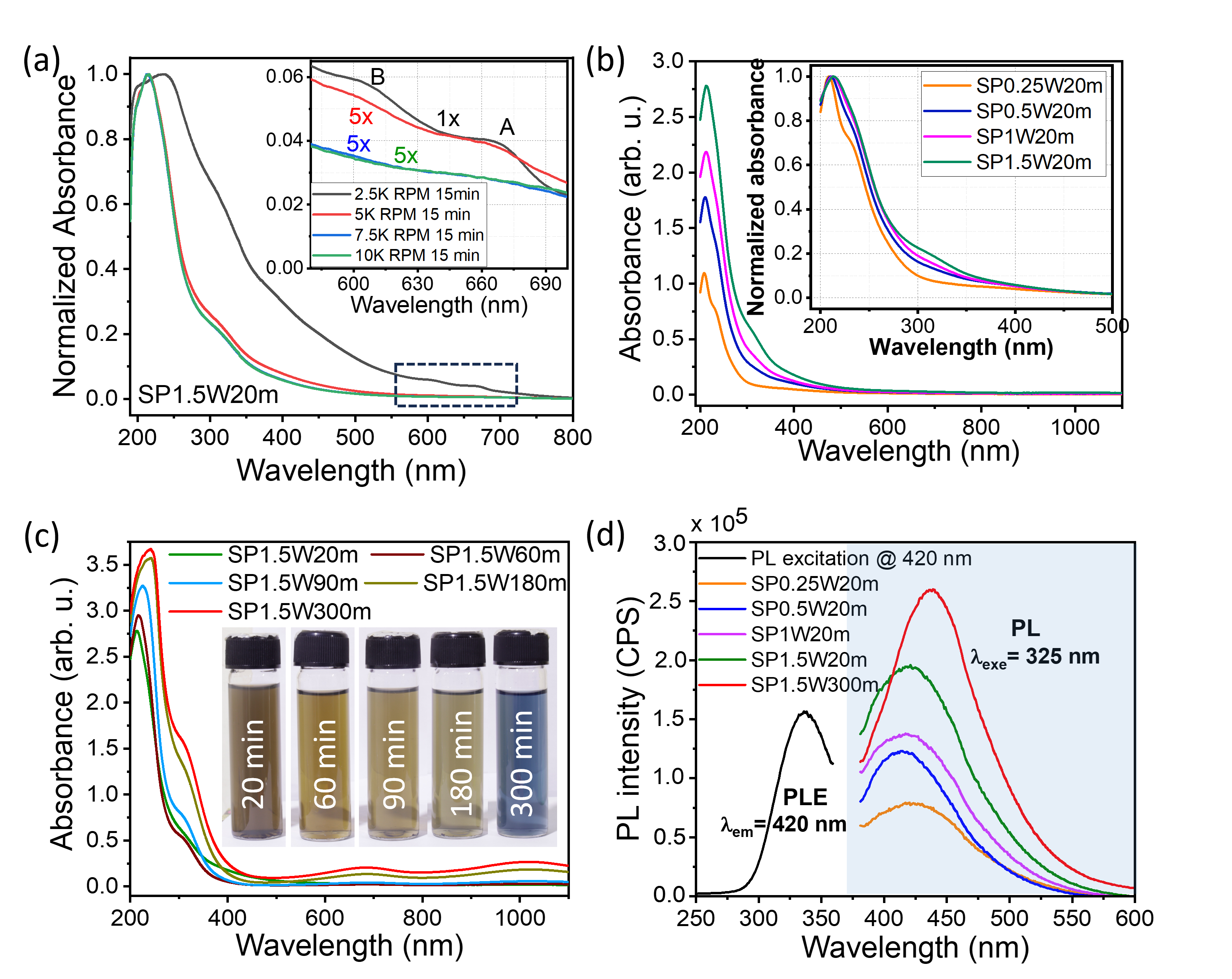}
 \caption{ The normalized UV-visible absorption for SP1.5W20m sample at different centrifuge speeds is shown in panel (a). The inset represents the magnified absorbance in the 550-700 nm range. UV-visible absorption spectra for samples synthesized at ablation powers ranging from 0.25 to 1.5 W are displayed in panel (b). The inset represents the normalized UV spectra. Panel (c) shows the absorption spectra for the samples prepared by increasing the ablation time from 20-300 min at 1.5 W ablation power. The camera images of the as-prepared samples are displayed in the inset. In panel (d), the PL emission spectra are depicted for samples excited with 325 nm excitation wavelength, and PL excitation (PLE) spectrum for SPI.5W20m  observed at 420 nm is also depicted. }
\label{fgr:P1}
\end{figure}

 UV-visible absorption spectra of the dispersed nanoparticle solution were used to determine the optical transitions and bandgap. The samples were centrifuged at various revolutions per minute (rpm) to better understand the aggregation impact and size distribution. Figure \ref{fgr:P1}(a) shows the absorption spectrum of the sample prepared at 1.5 W and 20 minutes of ablation time (SP1.5W20m). After the ablation, the stock solution was centrifuged at 2,500-10,000 rpm for 15 minutes, and the supernatant was meticulously extracted using a micropipette. Initially, at 2,500 rpm, characteristic  $\mathrm{A}$ and $\mathrm{B}$ exciton peaks were detectable at 674 nm (1.84 eV) and 613 nm (2.02 eV) from suspension micro- or nano-flakes. \cite{SUNITHA2018}.  However, the strong peak for the UV regime indicates quantum confinement of $\mathrm{MoS_{2}}$ QDs. The exciton peaks were not observed when the centrifuge speed was raised from 5,000 to 10,000 rpm, indicating the elimination of micro-flakes (cf. inset Figure 1 (a)). Considering particle size homogeneity for QDs and to study the confinement effect in the UV regime, samples centrifuged at 10,000 rpm were chosen for further study.\\
 
The absorption spectra of samples synthesized at different average powers at a constant ablation time of 20 minutes are illustrated in detail in Figure \ref{fgr:P1}(b). The sample prepared by varying ablation power at 0.25 W, 0.5 W, 1 W and 1.5 W are referred to as $\mathrm{SP0.25W20m}$, $\mathrm{SP0.5W20m}$, $\mathrm{SP1W20m}$ and $\mathrm{SP1.5W20m}$, respectively. All samples exhibited a UV-regime absorption peak at 215 nm with a shoulder peak at 310 nm that grew gradually with ablation power, which was prominent for $\mathrm{SP1W20m}$ and $\mathrm{SP1.5W20m}$ samples. Also, the ablation time can influence the optical properties of laser-ablated $\mathrm{MoS_{2}}$ QDs significantly. For a fixed power of 1.5 W, the samples were prepared by varying the ablation time from 20-300 minutes systematically, which are named as $\mathrm{SP1.5W20m}$, $\mathrm{SP1.5W60m}$, $\mathrm{SP1.5W90m}$, $\mathrm{SP1.5W180m}$, and $\mathrm{SP1.5W300m}$. The absorption spectra of these samples are depicted in Figure \ref{fgr:P1} (c), representing the systematic change in both UV and NIR regimes with the variation of sample preparation time. Notably, as the ablation time increased, the color of the colloidal solution changed from pale yellow to blue, as illustrated in Figure \ref{fgr:P1}(c) inset. Also, shoulder peak intensity was found to be increasing substantially with ablation time, along with two peaks in the NIR regime, which are different from the characteristic $\mathrm{A}$ and $\mathrm{B}$ peaks of $\mathrm{MoS_{2}}$. These observed peaks are centered at 680 nm and 1018 nm (cf. SI Figure S4), which are reported as the absorption peaks caused by surface plasmon resonance in $\mathrm{MoO_{3-x}}$ \cite{Alsaif2014,Annu2020,ZAMORAROMERO2020,Li_2021}. The surface of nanoparticles that are exposed to laser pulses for a longer time can lead to the formation of these oxide states. Increased ablation time promotes the oxidation of $\mathrm{MoS_{2}}$ QDs' surface, resulting in the formation of NIR absorbance peaks and a blue-colored colloidal solution.\\

In PL spectra, unlike the bulk features of $\mathrm{MoS_{2}}$ \cite{Eda2011}, only blue emission was observed in laser-ablated $\mathrm{MoS_{2}}$ QDs \cite{Doolen1998,Gan_2015,Gopalakrishnan2015,NGUYEN2019,Sung2020}. The PL emission spectra measured using an excitation wavelength of 325 nm are shown in Figure \ref{fgr:P1} (d). The spectral data exhibit a wide emission spectrum spanning from 380 to 500 nm, with the peak emission centered at about 420 nm. With an increase in the ablation power during the sample synthesis, there is a noticeable enhancement in the PL emission. This enhancement appears to be associated with the emergence of the tail state, observed around 286 nm in the absorption spectra, as depicted in the inset of Figure \ref{fgr:P1} (b). Also, there is a redshift in the PL peak position for S1.5W20m to SP1.5W300m, which is observed for the intermediate ablation-time dependent samples, as mentioned in SI figure S6. Notably, the PL excitation (PLE) spectrum for SP1.5W20m was detected at 335 nm for 420 nm emission, indicating a Stokes shift of  95 nm ($\sim$ 0.75 eV). This finding highlights the need for further investigation. 


\begin{figure}[h!]
\renewcommand{\thefigure}{2}
 \centering
\includegraphics[width=\linewidth]{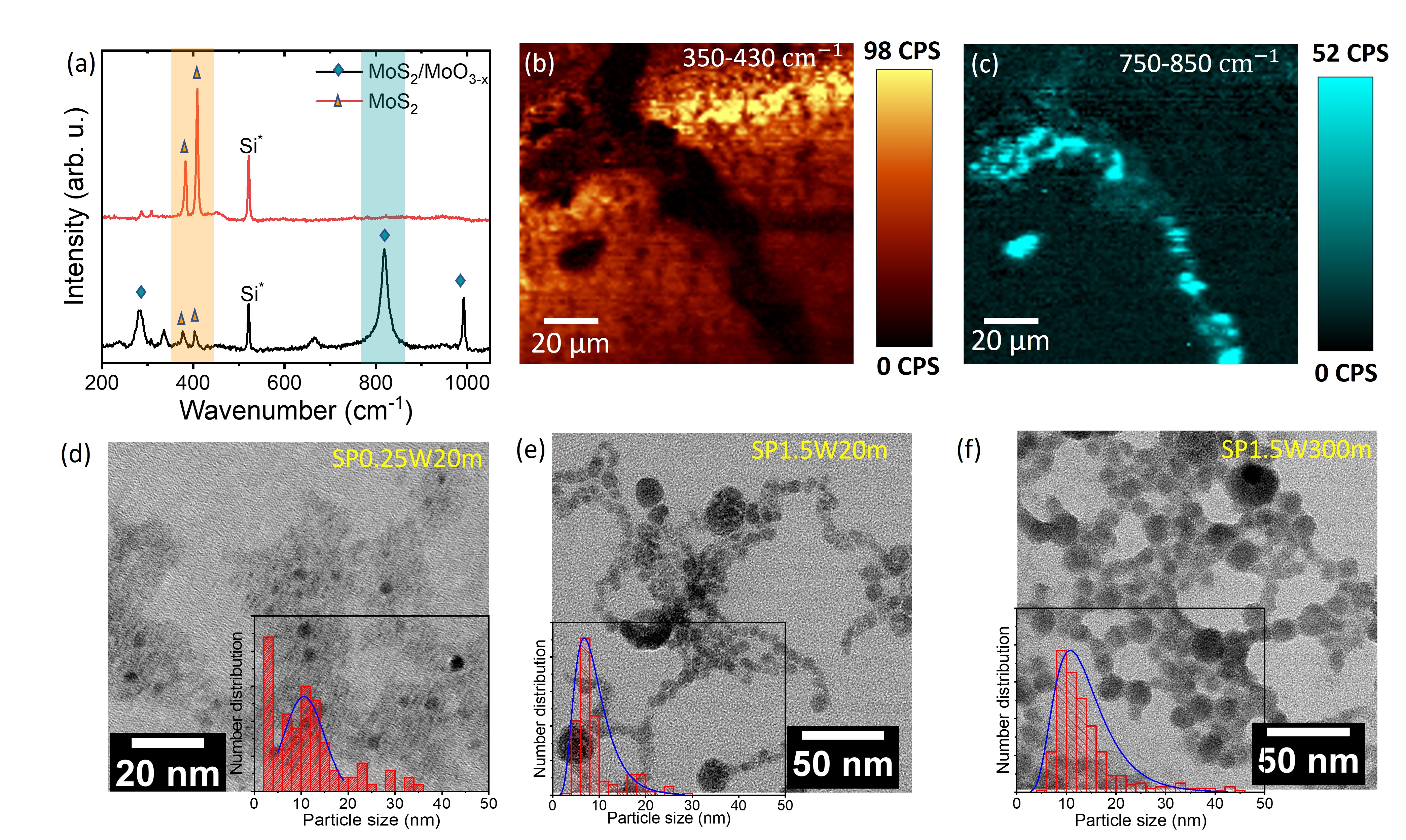}
 \caption{(a) Raman spectra for $\mathrm{MoS_{2}}$ and $\mathrm{MoS_{2}}$/$\mathrm{MoO_{3-x}}$ QDs are compared, respectively. The $\mathrm{E_{2g}^{1}}$ and $\mathrm{A_{2g}}$ modes for $\mathrm{MoS_{2}}$ are shown by triangle, while the $\mathrm{MoO_{3}}$ modes in $\mathrm{MoS_{2}}$/$\mathrm{MoO_{3-x}}$ heterostructure are represented by diamond. Raman mapping for $\mathrm{SP1.5W300m}$ sample is depicted in panels (b) for $\mathrm{MoS_{2}}$ modes in the range of $\mathrm{350-430\ cm^{-1}}$ (Orange) and (c) for $\mathrm{MoO_{3}}$ modes in the range $\mathrm{750-1050 \ cm^{-1}}$ (sky). The TEM micrographs for samples $\mathrm{SP0.25W20m}$, $\mathrm{SP1.5W20m}$ and $\mathrm{SP1.5W300m}$ in panel (d-f) where inset pictures represent the size distribution histogram.}
\label{fgr:P2}
\end{figure} 

Formation of $\mathrm{MoS_{2}}$ along with $\mathrm{MoO_{3-x}}$ phase is observed with the help of Raman spectroscopy, shown in Figure \ref{fgr:P2} (a). The characteristic modes $\mathrm{E^{1}_{2g}}$ and $\mathrm{A_{1g}}$ for $\mathrm{MoS_{2}}$ are observed at 382 $\mathrm{cm^{-1}}$, and 408 $\mathrm{cm^{-1}}$, respectively in both cases. The difference of $\mathrm{26 \ cm^{-1}}$ between $\mathrm{E^{1}_{2g}}$ and $\mathrm{A_{1g}}$ modes indicates the multi-layered nature of $\mathrm{MoS_{2}}$ nanoparticles \cite{Hong2012,Gnanasekar2018}. Additional peaks, positioned at $285\ \mathrm{cm^{-1}}$, $820\ \mathrm{cm^{-1}}$  and $995\ \mathrm{cm^{-1}}$, correspond to $\mathrm{MoO_{3}}$ phase, found in the positions where  $\mathrm{MoS_{2}}$/$\mathrm{MoO_{3-x}}$ heterostructure is observed \cite{SANTOS2012,Li_2021}. As observed, there is a significant change in the intensity ratio of $\mathrm{A_{1g}}$ and $\mathrm{E^{1}_{2g}}$ 2.3 to $\sim 1$ and broadening in these peaks c.f. Figure S8 (SI). $\mathrm{A_{1g}}$ mode is the phonon vibration of sulfur (S) atoms in opposite directions along c-axis in a plane perpendicular to the basal plane, whereas $\mathrm{E^{1}_{2g}}$ mode is the vibration of molybdenum (Mo) and sulfur in opposite direction but in the basal plane. The presence of electron-phonon coupling (EPC) leads to significant change in the $\mathrm{A_{1g}}$ mode along with decrease in change in ratio \cite{Lee2019,ko_2017}. Again, the charge transfer mechanism in heterostructures induces local dipole moment, which also depends on the direction of bonds. The formation of surface $\mathrm{MoO_{3-x}}$ phase changes the Mo-S bond direction as well the effective electronic environment, thus reducing the peak intensities from pure one \cite{Chakraborty_PRL2012,G_Sharma2020}.It is a confirmation of formation of heterostructures. Raman mapping was also carried out to analyze the formation and distribution of heterostructures. The Raman-mapped regions of $\mathrm{MoS_{2}}$ and $\mathrm{MoO_{3-x}}$ phases are shown in panels (b) and (c), respectively, for $\mathrm{S1.5W300m}$ sample. In Figure \ref{fgr:P2} (c), the Raman mapping clearly illustrates the presence of $\mathrm{MoO_{3}}$, which confirms the formation of $\mathrm{MoS_{2}}$/$\mathrm{MoO_{3-x}}$ QDs in these laser-ablated samples. \\


The size distribution histograms of these QDs were analyzed using TEM micrographs, as shown in Figure \ref{fgr:P2} (d-f).
Notably, the size distribution was found to change with ablation conditions. $\mathrm{SP0.25W20m}$ sample has a sharp peak at 5 nm with a broad peak centered at 10 nm. For $\mathrm{SP1.5W20m}$ and $\mathrm{SP1.5W300m}$ samples, the size distribution changes from 5-50 nm, but the maximum number of particles falls in the range of 5-15 nm and 10-15 nm, respectively. The size distribution is centered at 7.5 nm and 10 nm for $ \mathrm{SP1.5W20m}$ and $\mathrm{SP1.5W300m}$ samples, respectively, where the distribution is broad for $\mathrm{SP1.5W300m}$. In all these cases, the majority of the particle size is less than 10 nm; these can be assigned as QDs (Bohr radius for $\mathrm{MoS_{2}}$ ($a_{0}$) $\sim2$ nm) \cite{Doolen1998,Gan_2015}. The UV spectra of $\mathrm{SP1.5W20m}$ and $\mathrm{SP1.5W300m}$ were deconvoluted, revealing a noticeable broadening and enhancement of the 286 nm peak intensity (cf. Figure S4). For the sample SP1.5W20m to SP1.5W300m, the PL emission demonstrates a peak shift from 420 nm to 440 nm. It shows that for SP1.5W20m, the stokes shift is 0.85 eV, and for SP1.5W300m, it is  1 eV.  The $\mathrm{SP1.5W300m}$ QDs exhibit a larger particle size and a wider size distribution, resulting in a red-shift in PL emission caused by the polydispersive effect.

\begin{figure}[h!]
\renewcommand{\thefigure}{3}
 \centering
\includegraphics[width=\linewidth]{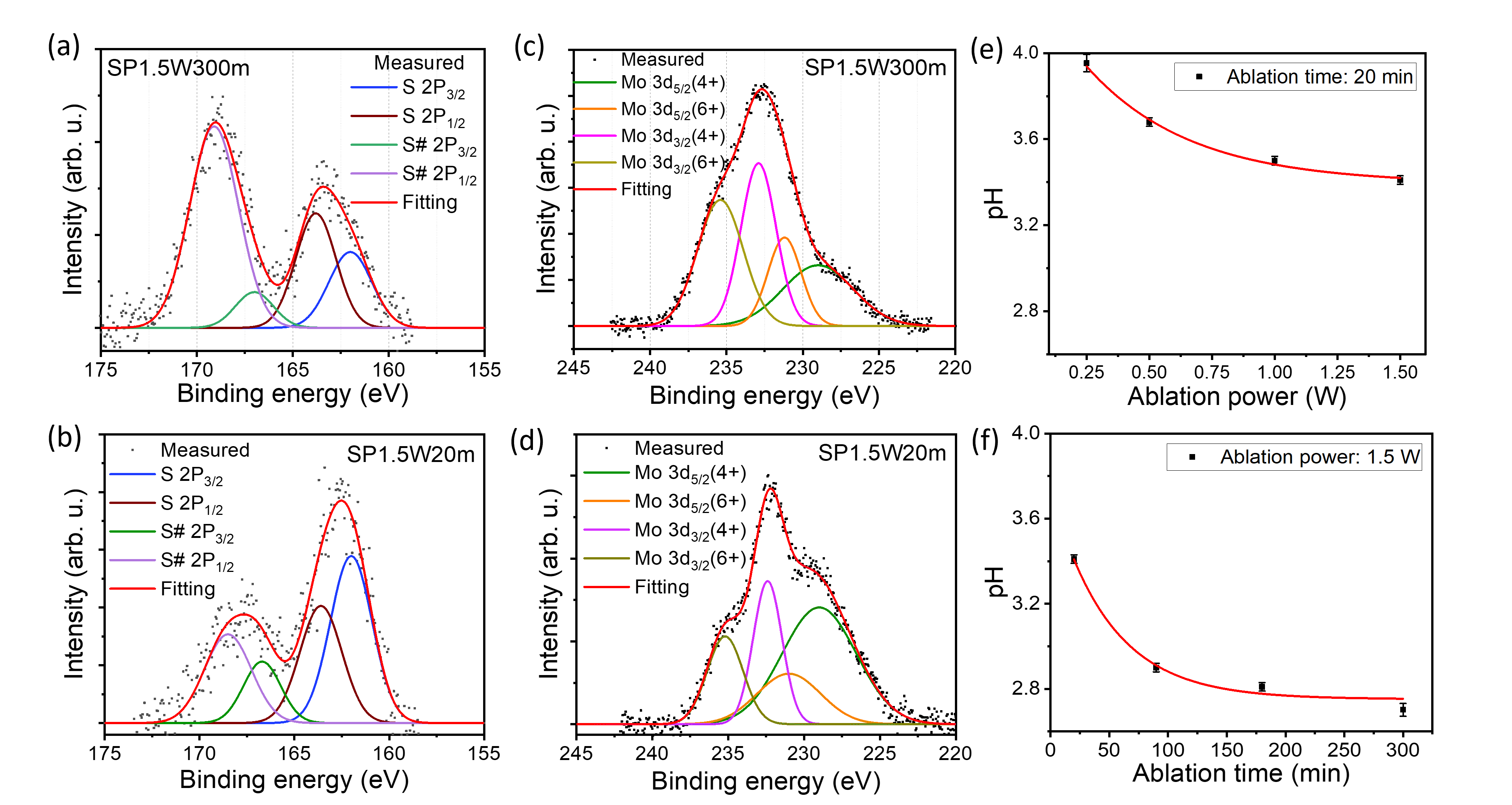}
 \caption{The XPS spectra of nanoparticles prepared for $\mathrm{SP1.5W20m}$ (a, c) and $\mathrm{SP1.5W300m}$ (b, d). The Mo 3d is plotted in panels (a) and (b), and S 2p is plotted in panels (c) and (d), respectively. The pH value of colloidal samples in solution form is measured and compared to samples made under conditions (e) by varying the ablation power and (f) ablation time.}
\label{fgr:P3}
\end{figure}

To confirm the oxidation states present in the chemical composition of particles, XPS was performed. The XPS data of Mo(3d) and S(2p) states for $\mathrm{SP1.5W20m}$ and $\mathrm{SP1.5W300m}$ samples are compared in Figure \ref{fgr:P3} (a-d). As reported, monolayer $\mathrm{MoS_{2}}$ XPS shows two sharp peaks at $\mathrm{231.9\ eV}$ and $\mathrm{228.8\ eV}$ for Mo(3d) $4^{+}$ oxidation state \cite{Prabhat_2020}. Also, S(2p) has two peaks, $\mathrm{162.9\ eV}$  and $\mathrm{161.7\ eV}$ corresponding to the S($\mathrm{2p_{1/2}}$) and S($\mathrm{2p_{3/2}}$). \cite{Afanasiev2019,Prabhat_2020}. The formation of $\mathrm{MoO_{3-x}}$ phase shows broadening in the Mo(3d) and S(2p) peaks due to the oxidation of both $\mathrm{Mo(4^{+}})$ and $\mathrm{S(2^{-}})$ species in $\mathrm{MoS_{2}}$ to higher oxidation states, i.e.  $\mathrm{Mo(5^{+})}$, $\mathrm{Mo(6^{+})}$, and $\mathrm{S(6^{-}})$ to form $\mathrm{MoS_{2-x}O_{x}}$  \cite{Xu_SR_2019,Afanasiev2019,Annu2020}.  Additional peaks at 232.2 eV and 235.4 eV are deconvoluted in panels (a) and (b) for $\mathrm{Mo(6^{+})}$, $\mathrm{3d_{5/2}}$ and $\mathrm{3d_{3/2}}$, respectively, indicating the presence of higher oxidation states \cite{Xu_SR_2019,Afanasiev2019}. 
A detailed tabulation showing the contribution of each species is depicted in the SI Table S2. It was found that $\mathrm{Mo(4^{+})}$ state has $\sim 72 \ \%$  and $\sim 53 \ \%$ contribution  for $\mathrm{SP1.5W20m}$ and, $\mathrm{SP1.5W300m}$ respectively; Accordingly, the contribution for higher oxidation state $\mathrm{Mo(6^{+}})$ increases from $ \sim 28 \ \%$ to $ \sim 47\ \%$ for them.  These findings are consistent with the contribution of the second band formation  in UV-visible ablation data. (c.f. SI Figure S4). Similarly, the S(2p) shows additional peak due to $\mathrm{S(6^{-}})$ state at $\mathrm{166.8\ eV}$ and $\mathrm{168\ eV}$, which is represented as $\mathrm{S^{\#}}$ in Figure \ref{fgr:P3} (c) and (d). The $\mathrm{S^{\#}}$ state represents the contribution of sulfate state (free sulfur species) in the sample \cite{Afanasiev2019}. These features confirm the contribution of higher oxidation states in $\mathrm{SP1.5W300m}$ samples. Moreover, the formation of free sulfur species was confirmed with the help of pH study of the colloidal solution, as shown in Figure \ref{fgr:P3} (e) and (f). The acidic nature of these samples confirms the formation of $\mathrm{H_{2}SO_{4}}$, which was also noticed in the XPS data as the formation of surface $\mathrm{SO^{2-}_{4}}$ radicals\cite{Afanasiev2019}. \\


The PL spectra for these QDs are the subject of discussion due to the large Stokes shift in the emission peak with respect to the excitation wavelength and absorption band edge. In the previous reports on fs-laser ablated $\mathrm{MoS_{2}}$ QDs, the PL emission was demonstrated at 400-450 nm. Most likely, the bands that are formed in the UV region are responsible for these emissions. However, a detailed understanding of these bands is required to know the mechanism of the optical transitions. Gan \textit{et al.} \cite{Gan_2015} reported that the blue emission of $\mathrm{MoS_{2}}$ QDs is a result of quantum confinement effect where the $\mathrm{A}$ and $\mathrm{B}$ exciton modes get blue shifted towards UV regime due to reduced particle size \cite{Chikan2002}. Nguyen \textit{et al.} \cite{NGUYEN2019} subsequently synthesized $\mathrm{MoS_{2}}$ nanoparticles using the fs-PLAL technique and examined the PL spectra based on surface states. The PL emission spectra were centered at around 420 nm for excitation wavelengths between 300 and 400 nm. However, the position of the emission peak remained unchanged.\\

In short, there are two possible reasons that are responsible for the blue emission: a) quantum confinement and b) surface state as the emission center. If blue emission is exclusively due to particle size reduction-induced quantum confinement, the peak shift in PL emission should be a wavelength-dependent process with no significant Stokes shift of 0.8-1.2 eV. Therefore, the surface state which could have formed below the conduction band must be thoroughly investigated. Tauc plot (shown in Figure S3, SI) confirms the formation of two individual bands in the UV regime, with one band edge in UV-C at 4.3 eV ($\mathrm{MoS_{2}}$) and a second one at UV-A 3.4 eV ($\mathrm{MoO_{3-x}}$) for SP1.5W300m. The band at 3.4 eV (360 nm) serves as the emission center and is potentially responsible for PL emission at 420 nm. As seen in Figure \ref{fgr:P1} (b), the PL intensity increases with the absorption intensity of the second band (c). We have confirmed these states as the higher oxidized state of $\mathrm{MoS_{2}}$ which forms $\mathrm{MoO_{3-x}}$ band in these QDs.\\

\begin{figure}[h!]
\renewcommand{\thefigure}{4}
 \centering
\includegraphics[width=\linewidth]{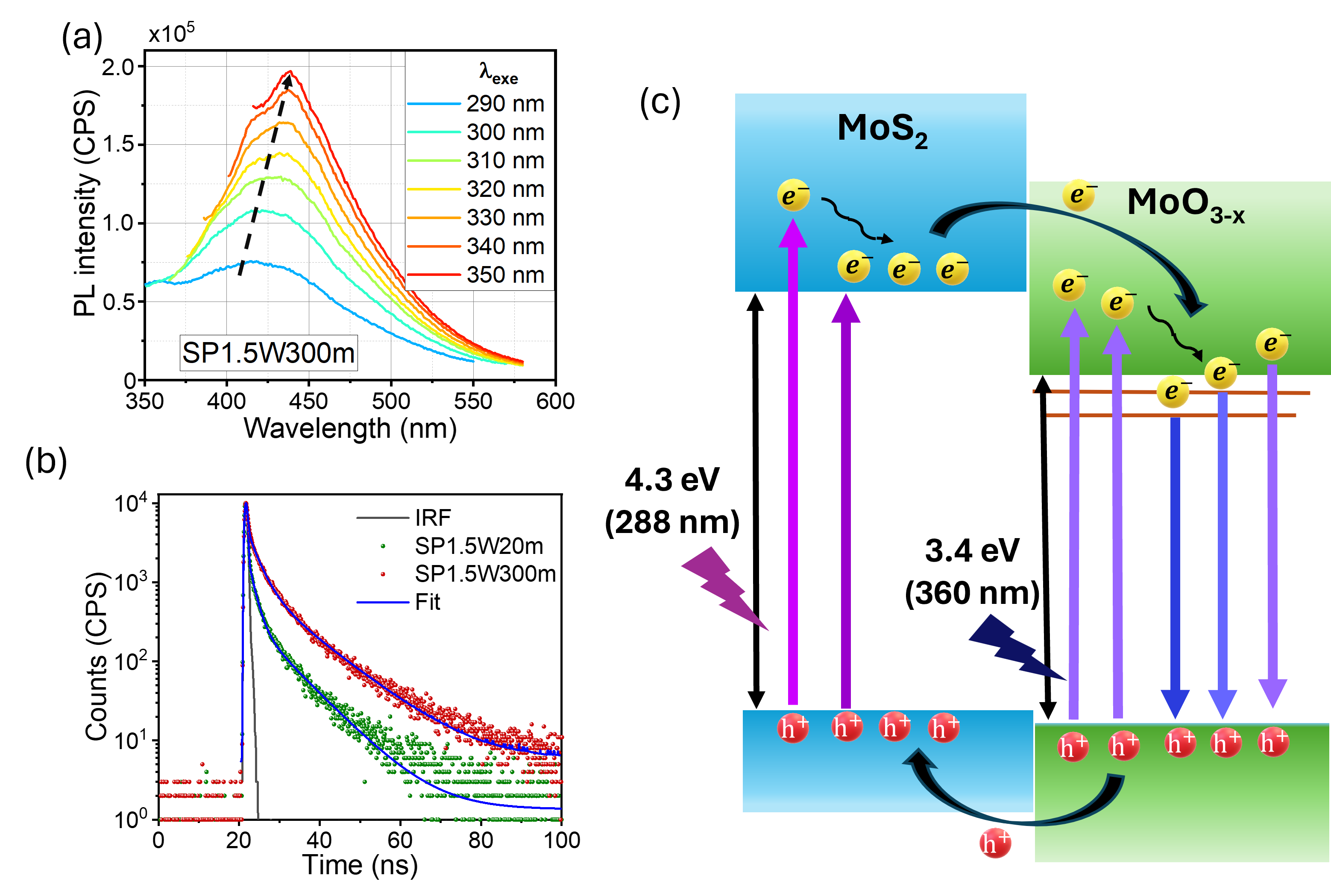}
\caption{ Panel (a) depicts the emission spectra for excitation wavelength-dependent PL for the $\mathrm{SP1.5W300m}$ sample. The TCSPC nanosecond transient-PL lifetime for $\mathrm{SP1.5W20m}$ and $\mathrm{SP1.5W300m}$ samples are compared in panel (b) for excitation at 340 nm and emission at 425 nm wavelength. In panel (c), the schematic for the electron transfer mechanism, where the inter-band transition is involved in the PL emission, is shown. }
\label{fgr:P4}
\end{figure}

Further, to have a clear understanding of Stokes shift, excitation wavelength-dependent PL was performed. The wavelength was varied in the range 290-350 nm in steps of 10 nm, as shown in Figure \ref{fgr:P4} (a) for $\mathrm{SP1.5W300m}$ sample. With increasing excitation wavelength, the PL emission intensity increases, predicting a larger emission yield around 350 nm. In addition, the steady redshift of the emission peak position as the excitation wavelength increases, is a strong signature of Stokes shift. Several research groups have tried to explain the origin of excitation-dependent luminescence on the basis of polydispersity of $\mathrm{MoS_{2}}$ QDs \cite{Hariharan2016,Ha2014,Wang_AnlChem_2014}. The redshift in the emission spectra indicates the quantum confinement effect in the QDs. Compared to $\mathrm{S1.5W20m}$ sample, $\mathrm{S1.5W300m}$ exhibits broader size distribution QDs \cite{Gan_2015}. Hence, higher excitation energies (290 nm) could excite smaller nanoparticles, whereas relatively low energy photons (330-350 nm) can excite comparatively larger sized nanoparticles, resulting in increased emission intensities at these wavelengths. At the same time, the difference between the excitation and emission wavelengths varies in the range of 0.8-1.2 eV for the corresponding 290-350 nm excitations.  The charge transfer suits best for the explanation since the excited electrons from $\mathrm{MoS_{2}}$ transfer charge non-radiatively to $\mathrm{MoO_{3-x}}$, that leads to blue emission with large Stokes shift. \cite{Peng_2016,Santiago2020,Liu_2021}.\\

Using nanosecond time-correlated single-photon counting (TCSPC), the transient-PL measurement was investigated to figure out how the intermediate transitions change during the decay process \cite{Doolen1998,Zhao2017,Wang2018,Bhattacharya_2020}. In this study, PL lifetime for 425 nm emission was measured using 340 nm excitation wavelength. The spectra show a slower lifetime for $\mathrm{S1.5W300m}$ when compared to $\mathrm{S1.5W20m}$, as illustrated in Figure \ref{fgr:P4}(b). These can be quantified by deconvolution of the transient-PL spectra using a biexponential decay function, where the $\tau_{1}$ and $\tau_{2}$ are assigned to fast and slow components, respectively. The $\tau_{1}$ was found to be $\mathrm{2.5\ ns}$ and $\mathrm{4\ ns}$, whereas $\tau_{2}$ was $\mathrm{9.7\ ns}$ and $\mathrm{13.2\ ns}$ for $\mathrm{S1.5W20m}$ and $\mathrm{S1.5W300m}$, respectively. The faster component is reported as radiative decay of charge carriers, which is the recombination of excited state carriers. The slower component can be attributed to trap or defect states, which can result from both radiative and non-radiative decay in the presence of defect states \cite{NGUYEN2015,NGUYEN2019,Bhattacharya_2020}. The slower decay process of $\tau_{2}$ for $\mathrm{S1.5W300m}$ indicates the formation of oxide states in these QDs, which are responsible for slowing down PL emission. The increase in lifetime is attributed to trap and defect-assisted delays in nonradiative recombination \cite{Tanoh2020}.  Additionally, the presence of oxygen vacancy levels below the conduction band (donor) creates more radiative channels, leading to enhanced PL yield and increased lifetime.   \\

\section{Conclusion}

 The overall picture for the study clearly indicates the formation of $\mathrm{MoO_{3-x}}$ phase in the laser ablated nanoparticles by this fs-PLAL method. The presence of these heterostructures within the nanoparticle systems has been verified through the analysis of Raman spectra, specifically by examining the characteristic peaks, $\mathrm{E^{1}_{2g}}$ and $\mathrm{A_{1g}}$ modes. The decrease in the intensity ratio of Raman of $\mathrm{A_{1g}}$ to $\mathrm{E^{1}_{2g}}$ and broadening in the full-width at half maximum indicates charge transfer from $\mathrm{MoO_{3-x}}$ to $\mathrm{MoS_{2}}$ phase \cite{Chakraborty_PRL2012,Li_Ptype2020}. Furthermore, the analysis of XPS spectra verifies the existence of a higher oxidation state as $\mathrm{MoO_{3-x}}$ phase. This phase shows a considerable rise from approximately $\sim 28 \%$  to approximately $\sim 47\ \%$ when comparing SP1.5W20m to SP1.5W300m. Furthermore, for SP1.5W20m, the deconvoluted absorption spectra for the secondary phase ($\mathrm{MoO_{3-x}}$) at the peak position at 286 nm indicate a contribution of $42 \% $, which is comparable with the results of XPS study.  The presence of this phase introduces more trap or surface states near the conduction band edge of the $\mathrm{MoO_{3-x}}$ phase. The increase in the fluorescence lifetime is an outcome of trap-defect assisted delay in both nonradiative and radiative channels. In this $\mathrm{MoS_{2}}/\mathrm{MoO_{3-x}}$ heterostructure, the conduction band (CB) of the $\mathrm{MoO_{3-x}}$ forms below the CB of $\mathrm{MoS_{2}}$ state, as shown in Figure \ref{fgr:P4}(c) \cite{Santosh2016,Li_2021}. In this schematic, a possible mechanism is shown  for $\mathrm{MoS_{2}}/\mathrm{MoO_{3-x}}$ 
 heterostructure, where electrons are transferred from $\mathrm{MoS_{2}}$ to $\mathrm{MoO_{3-x}}$ and holes from $\mathrm{MoO_{3-x}}$ to $\mathrm{MoS_{2}}$ \cite{Mirabbos2019,Suneel2020} due to their type-II heterostructure band alignment\cite{Li_2021}. The photoexcited electrons in $\mathrm{MoS_{2}}$ relax to minima of CB of $\mathrm{MoS_{2}}$ before electrons transfer from  $\mathrm{MoS_{2}}$ to  $\mathrm{MoO_{3-x}}$. Similarly, at relatively low energy photoexcitation, electrons get excited directly in the $\mathrm{MoO_{3-x}}$ phase.  Also, the oxygen vacancies in $\mathrm{MoO_{3-x}}$ phase lead to the dangling of Mo bonds, which form trap states below the CB of $\mathrm{MoO_{3-x}}$ \cite{Santosh2016}. The photoexcited electrons from $\mathrm{MoS_{2}}$ state get transferred to both the CB and trap states of $\mathrm{MoO_{3-x}}$ phase, which subsequently relaxes to the ground state, resulting in blue luminescence.

Conclusively, the overall optical properties of fs-PLAL synthesized QDs were found to be controlled by carefully choosing the laser parameters. It was found that the longer ablation time ($>$ 60 minutes) enhances the optical properties by forming an additional phase of $\mathrm{MoO_{3-x}}$ along with $\mathrm{MoS_{2}}$. Also, this hybrid structure is responsible for the blue emission and its large Stokes shift in PL study. The mechanism is proposed based on the electron transfer from $\mathrm{MoS_{2}}$ to $\mathrm{MoO_{3-x}}$ to elucidate the Stokes shift, which is also validated from PL lifetime measurements. These findings are applicable not only for laser-ablated $\mathrm{MoS_{2}}$ QDs systems but also for QDs of other 2D TMDCs synthesized in PLAL process for next-generation optoelectronics.
\begin{suppinfo}

\subsection{Preparation of $\mathrm{MoS_{2}}$ QDs using pulsed laser ablation in liquid}

\begin{figure}[h!]
 \centering
 \renewcommand{\figurename}{Figure S1}
 \renewcommand{\thefigure}{}
\includegraphics[width=\linewidth]{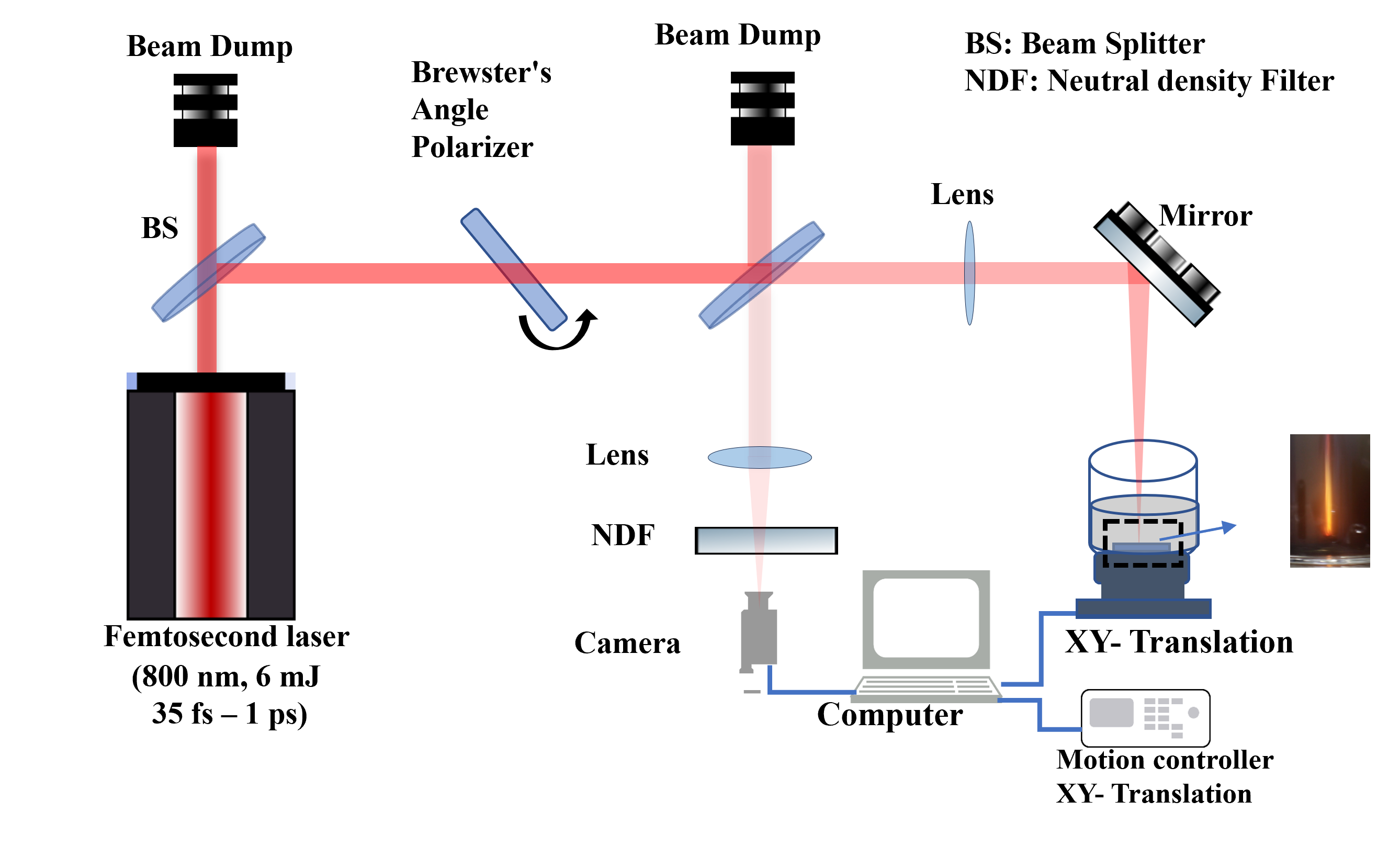}
 \caption{ The schematic of the experimental setup used for laser ablation is depicted in the figure. The Brewster’s angle polarizer was used for the variation of the input pulse energy, and a 20 cm lens was used for focusing the pulse on $\mathrm{MoS_{2}}$ pellet immersed in water medium. The digital image of the laser pulse ablating the pellet at 1.5 W and 500 fs pulsewidth is shown on the right.}
\label{fgr:setup}
\end{figure}

We employed a Ti: Sapphire femtosecond (fs) laser (Coherent, Astrella) operating at a repetition rate of 1 kHz and a central wavelength of 800 nm to create $\mathrm{MoS_{2}}$ quantum dots (QDs) through femtosecond pulsed laser ablation in liquid (fs-PLAL). An opaque target was made from $\mathrm{MoS_{2}}$ powder (Sigma-Aldrich) using a hot press pelletizer maintained at a temperature of $100\, ^\circ$C and pressure of 5 tonnes and then sintered at $200\, ^\circ$C for 12 hours. The ablation of this $\mathrm{MoS_{2}}$ target was carried out in deionized water (DI) water medium, and an optimized pulse width of 500 femtoseconds was also used in the process. A schematic representation of the fs-PLAL setup is shown in Figure S1. During ablation, the average laser power was also varied in the range of 0.25-1.5 W by selectively changing the Brewster angle polarizer. To have fresh interaction, the sample was moved continuously with the help of a motorized XY-translation stage. In this synthesis, the interaction of the laser beam with the target effectuates $\mathrm{MoS_{2}}$ nanomaterials of different size distributions in the DI water medium. We utilized 30 ml of deionized water in a 50 ml Borosil glass beaker, ensuring that the water level was maintained at a height of 3 cm from the target. A 20 cm focal length lens was used so that the Rayleigh range was completely immersed in water in order to avoid air-plasma interaction. The ablation time was varied from 20-300 min for the synthesis of $\mathrm{MoS_{2}}$ QDs. Post-ablation, the colloidal solutions were centrifuged at different rotation speeds (rotations per minute (rpm), 2,500-10,000 rpm) to remove undesired bulk particles. The supernatants of centrifuged samples are collected and stored in borosilicate glass bottles for characterization.

\subsection{Characterizations}
The UV-Vis-NIR absorption spectra of the prepared solutions were recorded with JASCO-V-570 UV-Vis absorption spectrophotometer in the wavelength range of 200-1100 nm. Photoluminescence (PL) emission spectra of the synthesized $\mathrm{MoS_{2}}$ QDs were characterized using a HORIBA Fluorolog-3 spectrofluorometer. Excitation wavelength-dependent PL emission studies were carried out by varying the excitation wavelength in the 280-400 nm range. Both absorption and PL study were carried out in a 1 cm quartz cuvette at room temperature in colloidal form. For excitation in PL study, the output of Xenon arc lamp mononochromaized via a monochromator, and the emission is collected on the photomultiplier tube after passing through a double stage monochromator, which is embedded in a spectrofluorometer.  During the measurement, the PL emission and PL excitation bandpass filter  were set to 5 nm and the measurement was carried out at step of 1 nm. The beam profile at the center of cuvette was having a top hat profile having size of $ \mathrm{12 mm \times 2 mm}$. The input beam has an energy of $ \mathrm{ 80\pm 5\ \mu J}$,  the fluence at cuvette is about $\mathrm{ 33\pm 2 \ mJcm^{-2}}$. For other studies, such as X-ray diffraction (XRD), Raman spectroscopy, and X-ray photoelectron spectroscopy (XPS) measurements, colloidal solutions were drop-cast on a $\mathrm{Si/SiO_{2}}$ substrate.\\

Raman spectra of the nanoparticles were investigated using a 532 nm diode laser in the range $200-1000\ \mathrm{cm^{-1}}$ using a WITec spectrometer (UHTS 300 VIS-NIR). The nanoparticles are drop casted on Si-substrate and Raman spectra measured using 50x microscope of numerical aperture (NA) = 0.55 (Zessis Epiplan-neofluar). The excitation laser has a central wavelength of 532 nm, whose spot size at focus is about 0.61 µm. For all measurement, the incident laser energy was maintained at  $\mathrm{ 0.5 \pm 0.05\ mJ}$ and corresponding the fluence was about $\mathrm{ 42 \pm 4 \ kJcm^{-2}}$. The fluence was kept below the  threshold value $~\mathrm{ 100\ kJcm^{-2}}$  where the phase transition from the $\mathrm{MoS_{2}}$ to $\mathrm{MoO_{3-x}}$ can not be possible \cite{Jagminas2016}. \\

For structural confirmation, an XRD pattern was recorded with a Rigaku Smart Lab diffractometer using $\mathrm{Cu\ k_{\alpha}}$ radiation ($\mathrm{1.54\ \AA}$). The particle size was analysed with the help of a transmission electron microscope (TEM), JEM-2100F from JEOL, with a focused beam at 200 keV. X-ray photoelectron spectroscopy (XPS) was carried out to investigate the oxidation state of laser-ablated nanoparticles using SPECS, GmbH (Germany). Al $\mathrm{K_{\alpha}}$ (1486.6 eV) was used as the excitation source, which was operated at $10 \ \mathrm{kV}$ and 10 mA current, and the analyzer chamber was maintained at $5 \times 10^{-10}$ mbar. Nanosecond transient PL was carried out using time-correlated single-photon counting (TCSPC) measurement with the help of the Flurocube lifetime system (Horiba JOBIN-VYON).   This is also done by using 1 cm quartz cuvette in colloidal form. The samples were excited using NanoLED-340 having a central wavelength of 340 nm and pulsewidth of 1 nm. The output pulse from NanoLED was focused using a 9 cm lens to the center of the cuvette. The pulse energy  was  $\mathrm{ 0.5 \pm 0.1\ pJ}$ , whose corresponding fluence at the focal volume was  $\mathrm{ 47\pm 9 \ mJcm^{-2}}$.   \\

\section{Optimization of centrifuge speed from UV-visible data}
The colloidal solutions obtained from fs-PLAL contain residues along with ablated nanoparticles. To get a uniform size distribution, the solution was centrifuged at different rotation speeds. The UV-visible absorption spectra for these nanoparticles collected after different rpm are compared in Figure S2.\\

\begin{figure}[h!]
\renewcommand{\figurename}{Figure S2}
 \renewcommand{\thefigure}{}
 \centering
\includegraphics[width=\linewidth]{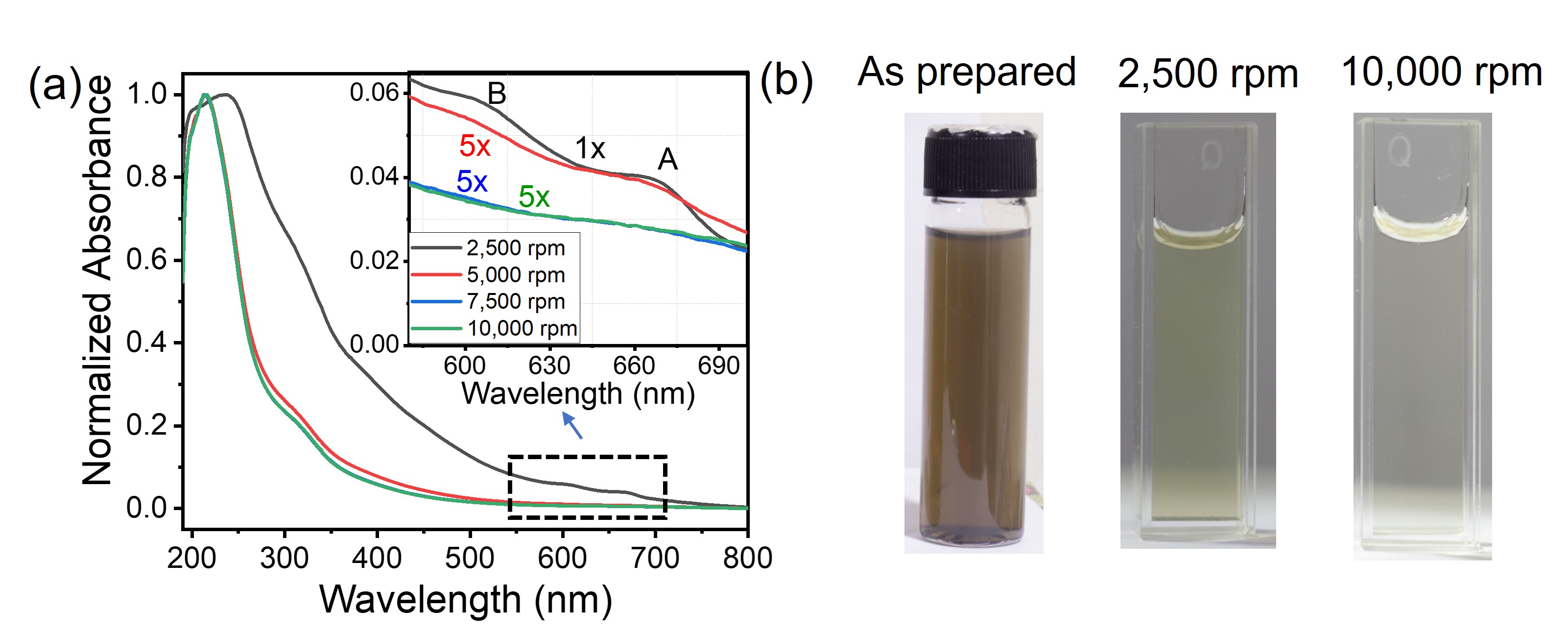}
\caption{The normalized UV-visible absorption is shown for a sample prepared at a 20 min ablation time and 1.5 W average power. The inset represents the magnified absorbance in the range of 550-700 nm. This represents the presence of A and B exciton features for multilayer $\mathrm{MoS_{2}}$ peaks for samples centrifuged at 2500 rotations per minute (rpm). These peaks are eventually eliminated from the 10,000 rpm-centrifuged sample. The digital image of the sample prepared at 2,500 and 10,000 rpm centrifuged sample at 1.5 W is shown in panel (b).}
\label{fgr:SP2}
\end{figure}

\newpage
\section{UV-visible absorption}
\begin{figure}[h!]
\renewcommand{\figurename}{Figure S3}
 \renewcommand{\thefigure}{}
 \centering
\includegraphics[width=\linewidth]{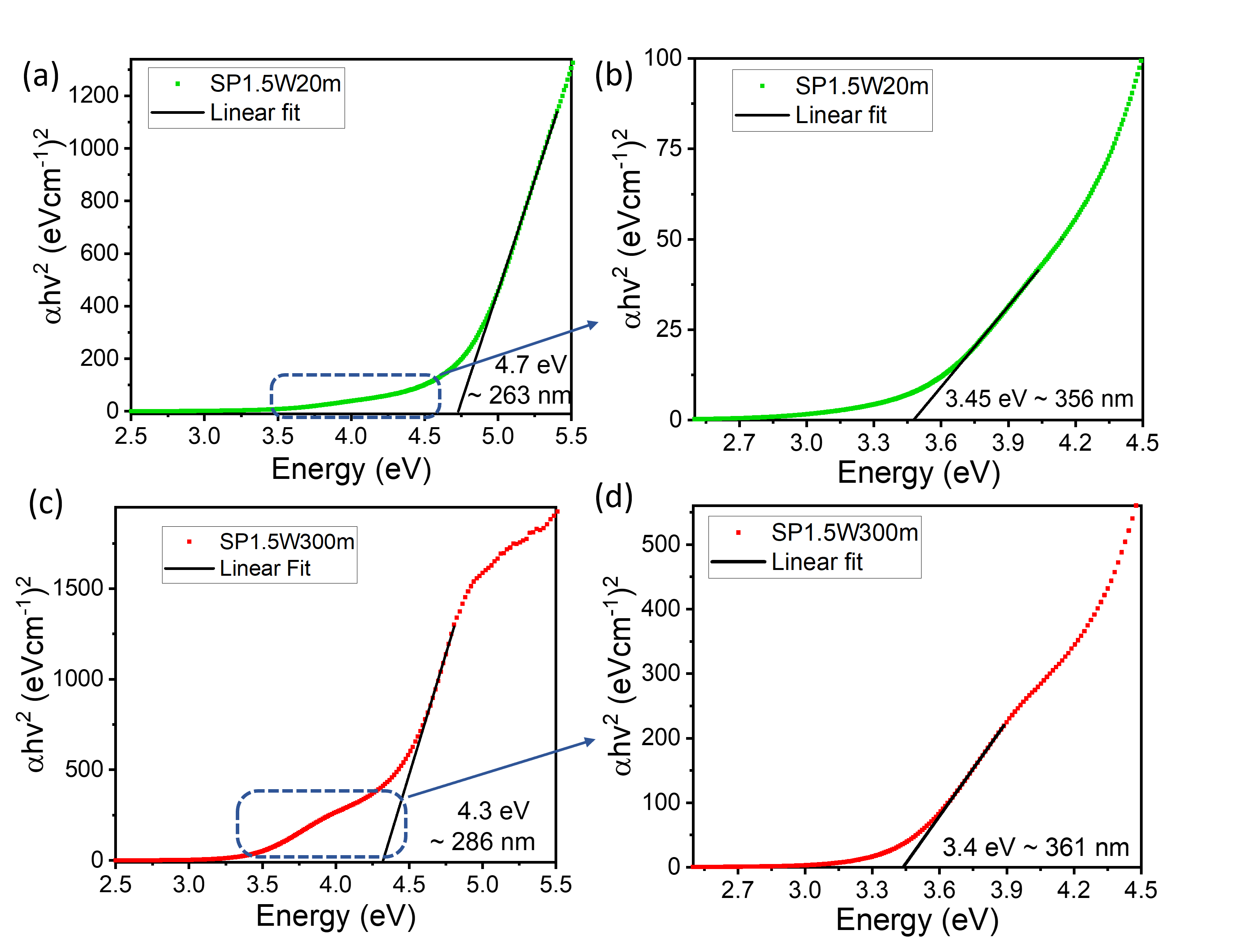}
 \caption{Tauc plots for samples SP1.5W20m and SP1.5W200m are shown in the figure above. These spectra have a strong absorption band in the UV-C regime corresponding to $\mathrm{MoS_{2}}$ QDs with a shoulder peak in the UV-A regime. The magnified images in the range of 2.6-4.5 eV are shown in panels b) and d) for their corresponding spectra in panels a) and b), respectively. The band at 3.4 eV shows a contribution of $\mathrm{MoO_{3-x}}$ band, which is more for SP1.5W300m sample compared to SP1.5W20m.}
\label{fgr:SP3}
\end{figure}

For a detailed understanding, samples prepared at 20 min and 300 min ablation times and 1.5 W average power are compared, which are named SP1.5W20 min and SP1.5W300 min, respectively. The Tauc plots of these samples are compared in Figure S3. In both cases, the UV-visible absorption spectra show two significant peaks in the UV regime. The first strong absorption edge is in the UV-C regime (4.3-4.7 eV), with the shoulder peak in the UV-A regime ($\sim$ 3.4 eV). The contribution of the UV-A band was found to be enhanced by increasing the ablation time. This peak was discussed as the surface state in the previous works \cite{Gan_2015}.
\\
Further, the absorption data was deconvoluted to multi-Gaussian fit to identify all the bands present in the absorption spectra. To compare the shift in peak position and the contribution of peak intensity, the peaks are normalized to the maximum intensity in the UV regime before fitting. The multi-Gaussian fitting is conducted in Origin using iterative least-squares nonlinear fitting of the experimental data. The complete absorption spectra were deconvoluted to 5 Gaussian with peak values fixed at (1) 213 nm, (2) (240-250) nm, (3) 286 nm, (4) 680 nm, and (5) 1018 nm. The deconvoluted spectra for samples dependent on ablation time are illustrated in Figure S4, showcasing (a) SP1.5W20m, (b) SP1.5W60m, (c) SP1.5W90m, (d) SP1.5W180m, and (e) SP1.W300m, sequentially. Peaks 1 and 2 correspond to the $\mathrm{MoS_{2}}$, which arises due to the quantum confinement of exciton peaks \cite{Chikan2002}, with peak 1 fixed at 213 nm and peak 2 allowed to vary between 240-250 nm, considering that an increase in particle size typically results in a redshift \cite{Gan_2015} Peak 3 represents the contribution of the secondary phase $\mathrm{MoO_{3-x}}$, observed at 286 nm \cite{Annu_ov__2020}. In addition to the peaks in the UV range, two near-infrared (NIR) peaks were detected, attributed to the localized surface plasmon resonance (LSPR). For fitting, these peaks were held constant at 680 nm and 1018 nm, denoted as peak 4 and peak 5, respectively. These peaks can be explained as localized surface plasmon resonance (SPR) in $\mathrm{MoO_{3-x}}$ phase of these QDs \cite{Annu2020,Li_2021}. The free electrons induced by oxygen vacancies in  $\mathrm{MoO_{3-x}}$ resonate to have absorption in the visible to NIR range \cite{Li_2021}. The two NIR absorption peaks originated from the absorption of the longitudinal and transverse plasmon modes from nonstoichiometric $\mathrm{MoO_{3-x}}$. This confirms the presence of $\mathrm{MoO_{3-x}}$ in laser-ablated QDs for a longer ablation time. A table containing all fitting parameters, including the area under the curve (A), fitted wavelength ($\lambda$) and FWHM ($\Delta \lambda$), is presented in Table S1. It is observed that the contribution of the 286 nm peak increases with ablation time, accounting for about $42\%$ for SP1.5W300m compared to $29\%$ for SP1.5W20m. The deconvolution of UV-visible absorption data in our case paves the way for understanding the $\mathrm{MoO_{3-x}}$ phase quantitatively. These results are also consistent with XPS analysis.

\begin{table}[h]
  \renewcommand\thetable{S1}
 \caption{The table indicates that the contribution of UV-visible peaks are calculated by deconvoluting into five Gaussian peaks.}
\label{table:transposed}
\begin{tabular}{|p{1.8cm}|p{2.4cm}|p{2.4cm}|p{2.4cm}|p{2.6cm}|p{2.6cm}|}
\hline
\textbf{Sample} & \textbf{SP1.5W20m} & \textbf{SP1.5W60m} & \textbf{SP1.5W90m} & \textbf{SP1.5W180m} & \textbf{SP1.5W300m} \\
\hline
$\mathrm{A_{1}}$ & 65 (71$\%$)  & 53 (68 $\%$) & 61 (60.3$\%$)  & 60 (45.2$\%$)  & 53 (38$\%$) \\
$\lambda_{1}$ (nm) & 213 & 213 & 213  & 213 & 213 \\
$\Delta \lambda_{1}$ (nm) & 56 & 61  & 56  & 60  & 55 \\
\hline
$\mathrm{A_{2}}$ & 0  & 2.9 (3$\%$) & 6.4 (6.3$\%$)  & 8.2 (6.2$\%$) & 9 (6.4$\%$)  \\
$\lambda_{2} $ (nm)  & 0  & 241.9& 241.2  & 248  & 249 \\
$\Delta \lambda_{2}$ (nm)  & 0  & 27 & 25 & 25  & 25 \\
\hline
$\mathrm{A_{3}}$ & 27 (29$\%$)  & 21 (26.8$\%$)  & 28.7 (28.3$\%$) & 48 (36.2$\%$)  & 59 (42.2$\%$) \\
$\lambda_{3}$ (nm) & 286  & 286  & 286 & 286  & 286 \\
$\Delta \lambda_{3}$ (nm) & 88 & 89  & 91  & 97  & 103 \\
\hline
$\mathrm{A_{4}}$ & 0  & 0.56 (0.6$\%$) & 1.5 (1.4$\%$) & 4.6 (3.4$\%$)  & 5 (3.3$\%$) \\
$\lambda_{4}$ (nm)   & 0  & 680 & 680 & 680  & 680 \\
$\Delta \lambda_{4} $ (nm) & 0  & 132 & 146 & 148  & 136.7 \\
\hline
$\mathrm{A_{4}}$ & 0  & 1.3 (1.6$\%$) & 3.8 (3.7$\%$) & 12 (9$\%$)  & 14.2 (10.1$\%$) \\
$\lambda_{5}$ (nm)  & 0  & 1018 & 1018 & 1018  & 1018 \\
$\Delta \lambda_{5}$ (nm) & 0  & 195 & 228 & 241 & 239 \\
\hline
\end{tabular}
\end{table}

\begin{figure}[h!]
\renewcommand{\figurename}{Figure S4}
 \renewcommand{\thefigure}{}
 \centering
\includegraphics[width=\linewidth]{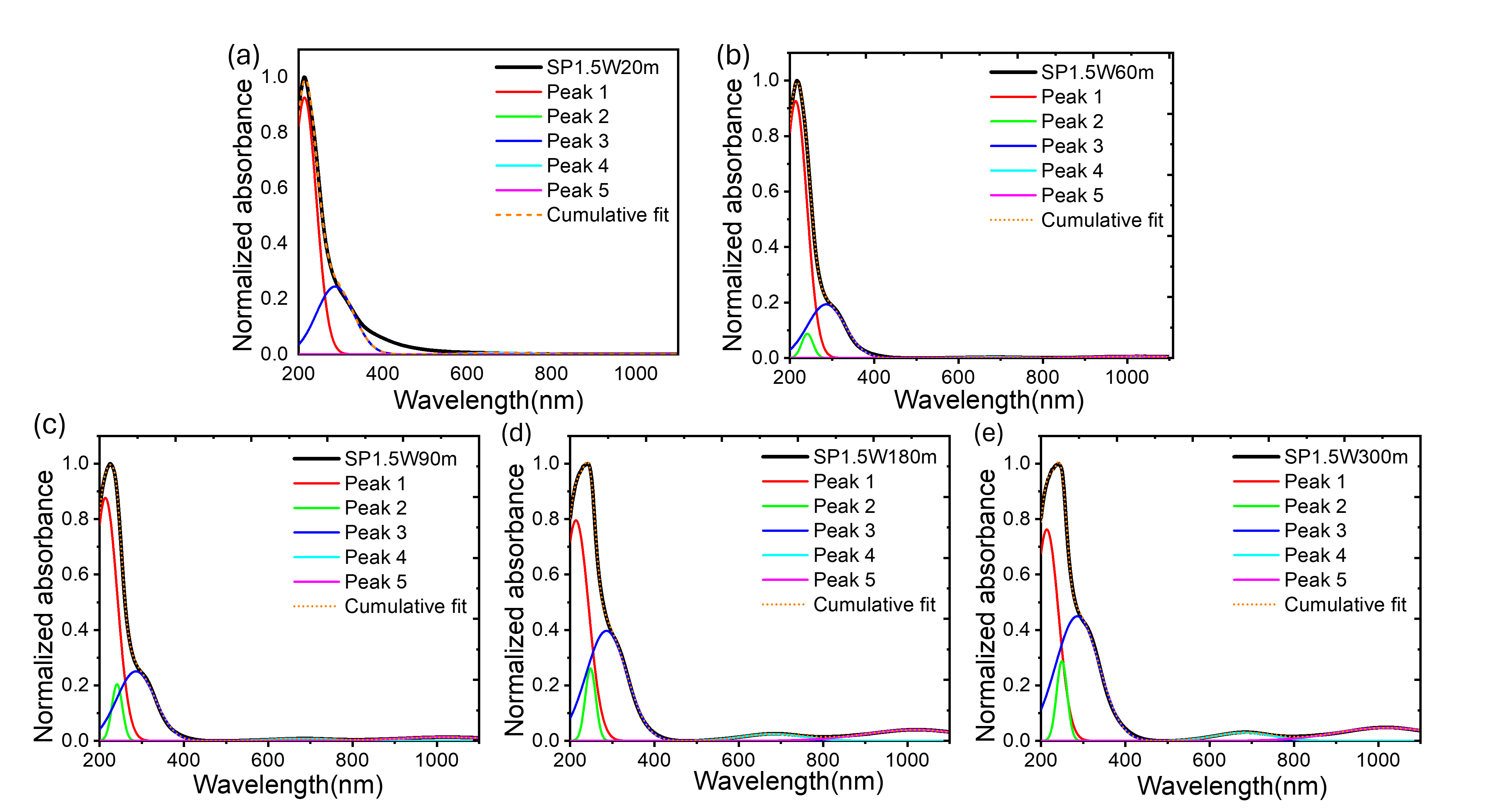}
 \caption{The deconvolution of UV-visible data are shown for (a) $\mathrm{SP1.5W20m}$, (b) $\mathrm{SP1.5W20m}$, (c) $\mathrm{SP1.5W20m}$ (d) $\mathrm{SP1.5W20m}$ and (e) $\mathrm{SP1.5W300m}$ samples.
}
\label{fgr:SP4}
\end{figure}

\newpage
To confirm these peaks, we have independently compared the commercial $\mathrm{MoO_{3}}$ powder to confirm these peaks. After the grinding, the commercial powder turned blue, as shown in Figure  S5.

\begin{figure}[h!]
\renewcommand{\figurename}{Figure S5}
\renewcommand{\thefigure}{}
 \centering
\includegraphics[width=\linewidth]{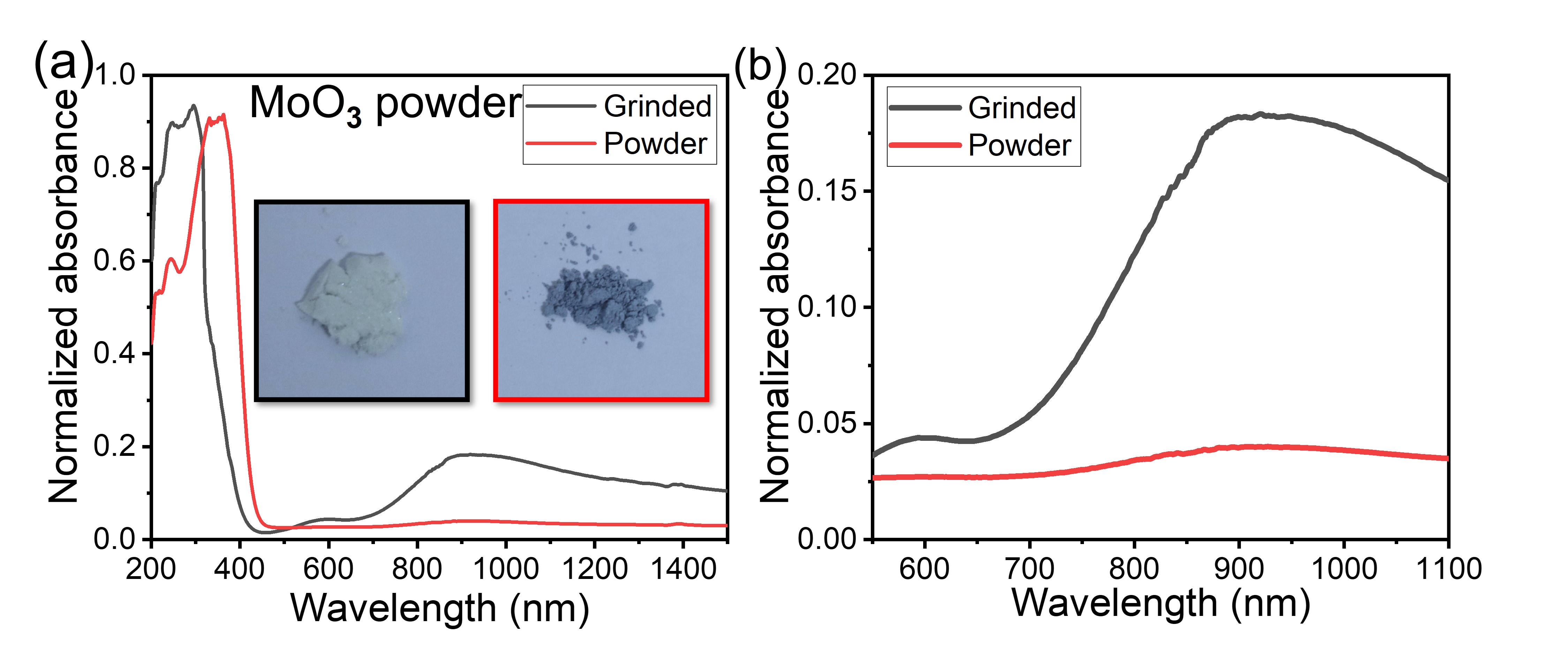}
 \caption{The normalized UV-visible absorption is depicted $\mathrm{MoO_{3}}$ in panel Figure S5. (a). The inset represents the powder that was used for taking the UV-visible absorption. The white powder is commercial $\mathrm{MoO_{3}}$ powder, and the blue-colored sample is the one that was ground for 30 min at room temperature. Panel b) shows the clear peaks in the 500-1100 nm range.
}
\label{fgr:SP5}
\end{figure}

\newpage
\section{PL emission and excitation study}

The PL emission and excitation spectra are measured in the colloidal phase using 1 cm cuvette. The PL emission spectra measured for samples synthesized by varying the ablation time is shown in Figure S6. As observed, the PL emission for the samples SP1.5W20m to SP1.5W300m which is more evident shows a gradual red shift in the PL spectra, which is even clearer in the normalized spectra c.f. Figure S6 (b). The redshift in the peak position can be related to the larger particle size distribution, whose mean position shifts from 7.5 nm to 11 nm for SP1.5W20m and SP1.5W300m.

\begin{figure}[h!]
\renewcommand{\figurename}{Figure S6}
\renewcommand{\thefigure}{}
 \centering
\includegraphics[width=1\linewidth]{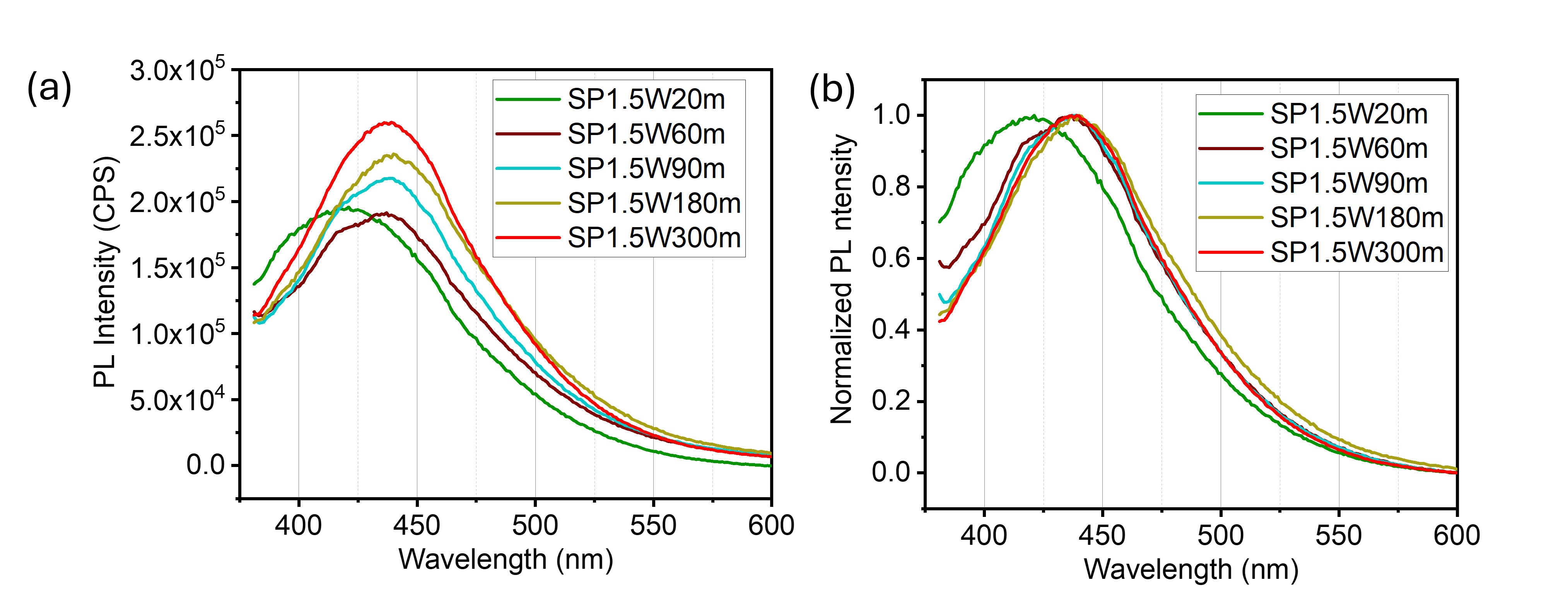}
 \caption{The PL emission spectra for samples varying prepared by varying ablation time are depicted in this figure for an excitation wavelength of 325 nm}
\label{fgr:SP6} 
\end{figure}

The PL excitation(PLE) measurement for 420 nm emission was carried out to probe the corresponding excitation band. In Figure S6 the PL excitation spectra for 420 nm emission are depicted. For 420 nm (2.95 eV), the excitation spectra show center at 335 nm (3.7 eV) which is a significant Stokes shift of about 0.75 eV.

\begin{figure}[h!]
\renewcommand{\figurename}{Figure S7}
\renewcommand{\thefigure}{}
 \centering
\includegraphics[width=0.6\linewidth]{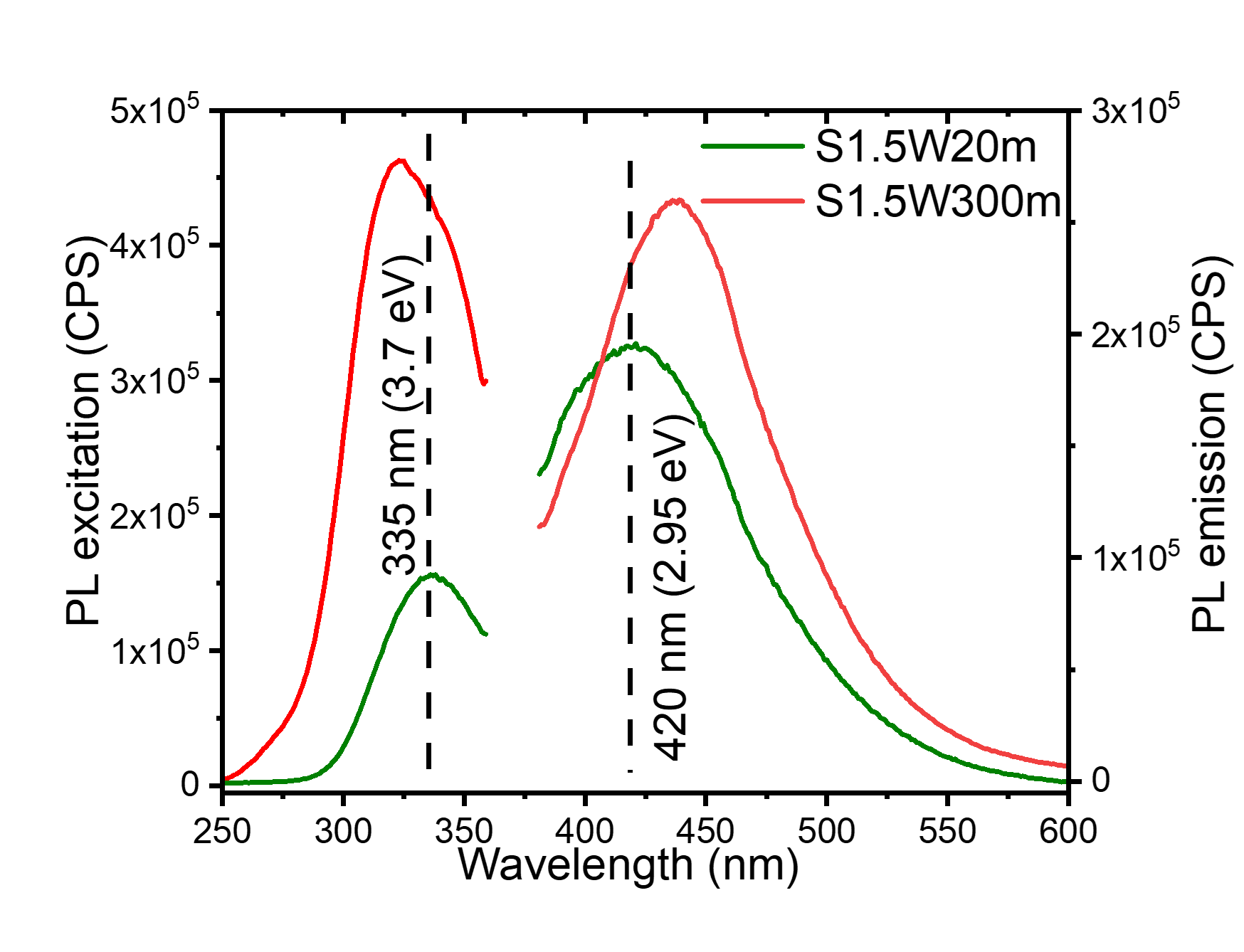}
 \caption{The PL excitation (PLE) spectra for 420 nm emission is shown for SP1.5W20m and SP1.5W300m samples. On the other axis (right) the PL emission at 325 nm is compared for them.  }
\label{fgr:SP7}
\end{figure}

\newpage
\section{Raman spectra}

The background shift was corrected by keeping the Si peak position at 520 $\mathrm{cm^{-1}}$. 
It is noteworthy that the peak positions exhibited alterations in the intensity ratio of $\mathrm{A_{1g}}$ and $\mathrm{E^{1}_{2g}}$, which were shown to decrease as the contribution of $\mathrm{MoO_{3}}$ increased, c.f. Figure S8(a). This was observed when specifically at 820 $\mathrm{cm^{-1}}$ and 960 $\mathrm{cm^{-1}}$ came to picture, which corresponds to $\mathrm{MoO_{3}}$.
To measure the variations in the Raman spectra, the characteristic peaks associated with the $\mathrm{MoS_{2}}$, $\mathrm{E^{1}_{2g}}$, and $\mathrm{A_{1g}}$ modes are fitted to Lorentzian peaks, as depicted in panels (b) and (c). As observed, there was a clear boarding of FWHM of peak ($\Delta \omega$) 3.3 to 6.5 $\mathrm{cm^{-1}}$  and 3.6 to 6 $\mathrm{cm^{-1}}$ for $\mathrm{E^{1}_{2g}}$ and $\mathrm{A_{1g}}$, respectively. In addition, the change in the intensity ratio of $\mathrm{I_{A_{1g}}/I_{E^{1}_{2g}}}$  changes from 2.6 to 1. \\

\begin{figure}[h!]
\renewcommand{\figurename}{Figure S8}
\renewcommand{\thefigure}{}
 \centering
\includegraphics[width=1\linewidth]{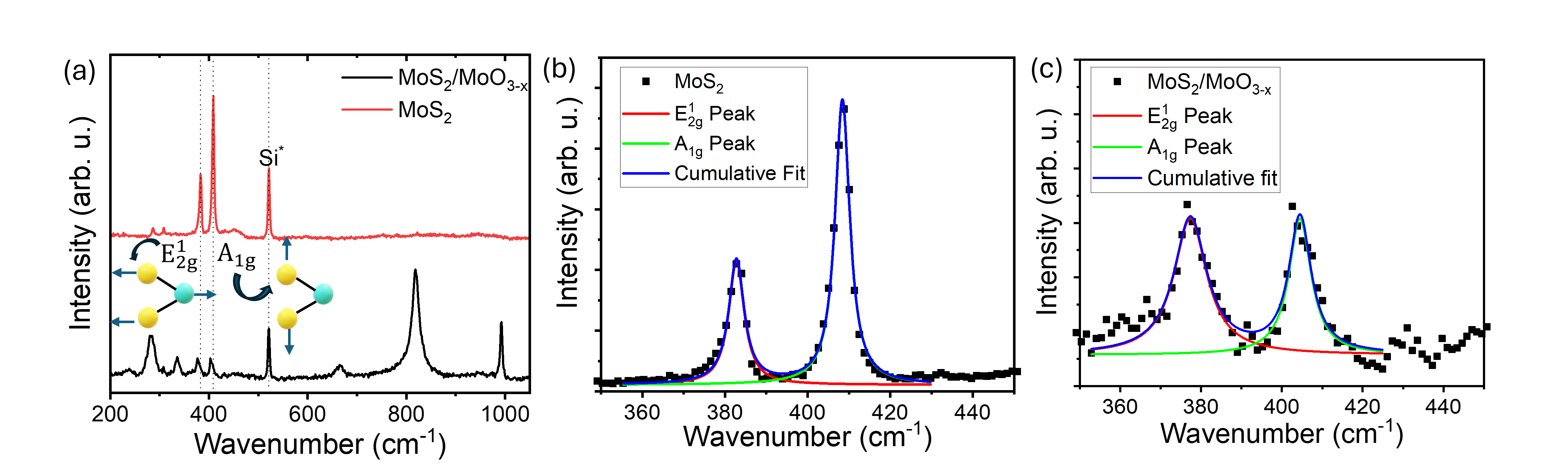}
 \caption{The Raman spectra for the different position are plotted showing different contribution of $\mathrm{MoS_{2}/MoO_{3-x}}$ phase in these QDs. The deconvolution of $\mathrm{E^{1}_{2g}}$ and $\mathrm{A_{1g}}$ for those points are shown in panel (b) and (c).}
\label{fgr:SP8}
\end{figure}

$\mathrm{A_{1g}}$ mode is the phonon vibration of sulfur (S) atoms in opposite directions along c-axis in a plane perpendicular to basal plane, whereas $\mathrm{E^{1}_{2g}}$ mode attribute to vibration of molybdenum (Mo) and sulfur in opposite direction but in the basal plane, the schematics are depicted in the inset of panel (b). The doping of an electron introduce lead to occupation in minimum of conduction band at K-point, which is characteristic of $(d_{z^{2}})$  Mo state \cite{Chakraborty_PRL2012}. The presence of electron-phonon coupling (EPC) leads to significant change in the A1g mode along decrease in change in $\mathrm{I_{A_{1g}}/I_{E^{1}_{2g}}}$ ratio \cite{Chakraborty_PRL2012,ko_2017} The FWHM broadening indicates lattice distortion due to formation of surface  $\mathrm{MoO_{3-x}}$ phases. In-plane tensile strain induces elongation of the lattice along the strain direction, leading to the redshift of $\mathrm{E^{1}_{2g}}$ modes for monolayer $\mathrm{MoS_{2}}$ \cite{Chakraborty_PRL2012,Lee2019}. It is also confirmation of the formation of heterostructure. The Raman peak intensities mostly explain the change in polarizability. Again, the charge transfer mechanism in HS induces local dipole moment, which also depends on the direction of bonds. In HS, the formation of surface  $\mathrm{MoO_{3-x}}$ phase changes the Mo-S bond direction as well the effective electronic environment, thus reducing the peak intensities from pure one. It is a confirmation of formation of heterostructure.

\newpage
\section{XPS study}
The deconvolution of XPS data was carried out using multi-Gaussian peak fit using Matlab. 
 The area under the curves is calculated to quantify the contribution of each peak, cf. Table S2. As calculated, the contribution of Mo 3d(6+) in SP1.5W300m is $47\%$ whereas $28\%$ for SP1.5W20m. This attributed to the presence of a higher oxidation phase in these nanoparticles, which is evidently due to the presence of $\mathrm{MoO_{3-x}}$ phase in $\mathrm{MoS_{2}}$ nanoparticles. Similarly, the contribution of Mo 3d (4+) attributed to $\mathrm{MoS_{2}}$  phase. As reported in this manuscript, the contribution of the $\mathrm{MoS_{2}}$  along with $\mathrm{MoO_{3-x}}$  has been observed with an increase in the laser ablation time. 
 
\begin{table}
\renewcommand\thetable{S2}
  \caption{The tabulation for all contributing species are presented in the XPS spectra are shown.} 
  \label{tbl_t2}
  \begin{center}
  \begin{tabular}{|l|l|l|l|ll|ll|}
    \hline
    \ & \textbf{Contributing}& \textbf{Species} &  \textbf{Binding}  & \multicolumn{2}{c|}{\textbf{SP1.5W20m}} & \multicolumn{2}{c|}{\textbf{SP1.5W300m}}\\ 
      & \textbf{Levels}& &  \textbf{Energy (eV)} & \multicolumn{2}{c|}{\textbf{Area (\%)}}&  \multicolumn{2}{c|}{\textbf{Area (\%)}}\\ 
   \hline
   \ Mo $(4^{+})$ & $\mathrm{MoS_{2}}$  & $3d_{5/2}$ & 229.1 & 60 & & 23 & \\
   \ Mo $(4^{+})$ & $\mathrm{MoS_{2}}$  & $3d_{3/2}$ & 232.3 & 12 (72\%) & & 30 (53\%) &\\
   \hline
   \ Mo $(6^{+})$ & $\mathrm{MoO_{3}}$& $3d_{5/2}$ & 231.3 & 16 & & 16 & \\
   \ Mo $(6^{+})$ &$\mathrm{MoO_{3}}$ & $3d_{3/2}$ & 235.4 & 12 (28\%) & & 31 (47\%) & \\ 
   \hline
   \ S $2p_{3/2}$ & $\mathrm{MoS_{2}}$ & $2p_{3/2}$ & 162 & 38.2 & & 17  & \\
   \ S $2p_{3/2}$ & $ \mathrm{MoS_{2}}$ & $2p_{1/2}$ & 163.2 & 26.5 (64.7\%) & & 24 (41\%) & \\
   \hline
   \ S $2p_{3/2}$ & Free Sulfur & $2p_{3/2}$ & 166.8 & 24.8 &  & 7.4 & \\
   \ S $2p_{1/2}$ & Free Sulfur & $2p_{1/2}$ & 168 & 10.5 (35.3\%) & & 51.6 (59\%) & \\
   \hline
  \end{tabular}
    \end{center}
\end{table}

\newpage
\section{pH measurement}
The $\mathrm{S^{2-}}$ species present in  Mo-S-Mo bonding undergoes a reaction with laser pulses and ambient solvent to form $\mathrm{S^{2-}_{2}}$. This further reacts with the OH group to form sulphuric acid \cite{Afanasiev2019}. These can be understood from the equation mentioned below \cite{Afanasiev2019} :

\begin{equation}
    \mathrm{MoS_{2}+14 H_{2}O+35O_{2}=4Mo_{2}O_{5}HSO_{4}+12H_{2}SO_{4}}
\end{equation}

The formation of additional sulphuric acid and substitution of oxygen for sulphur in the Mo coordination are responsible for an acidic pH. These reactions favour the longer exposure of laser pulses at higher ablation powers (c.f. Figure S9).

\begin{figure}[h!]
\renewcommand{\figurename}{Figure S9}
 \renewcommand{\thefigure}{}
 \centering
\includegraphics[width=\linewidth]{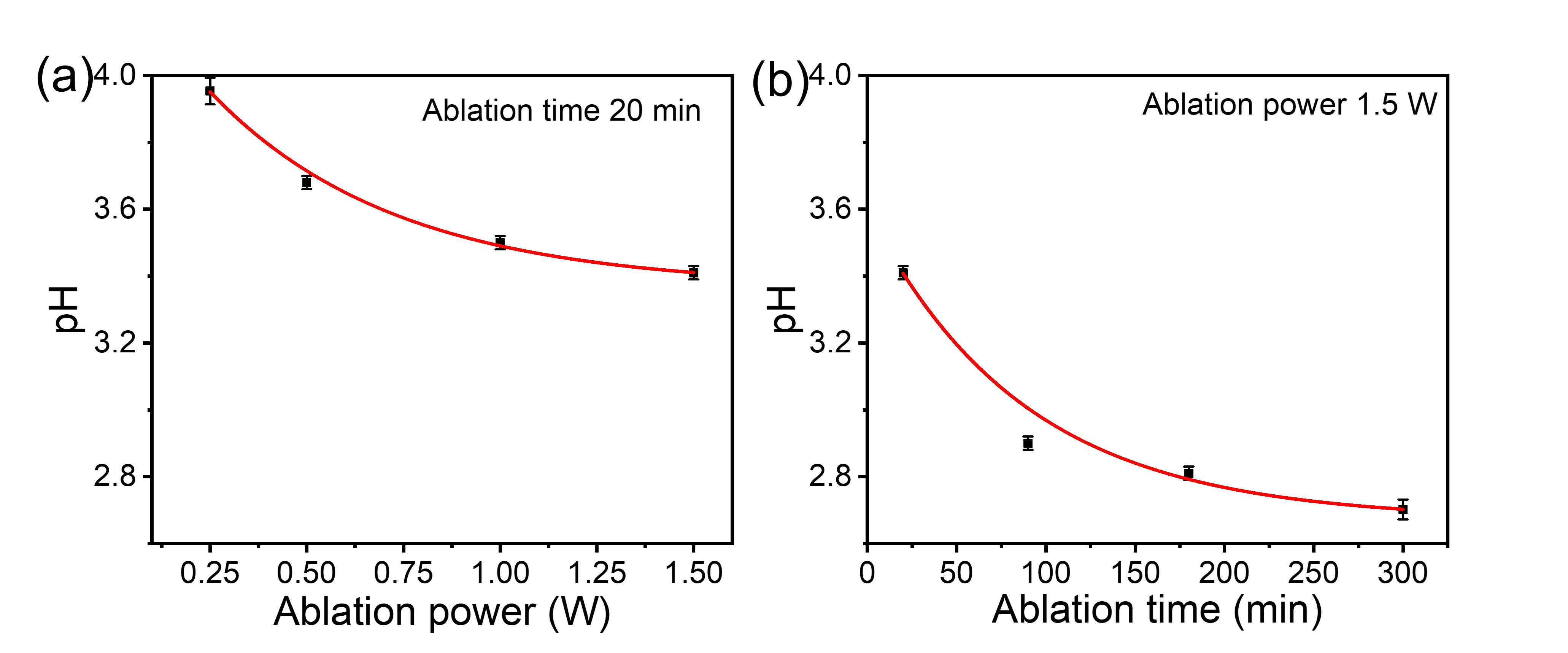}
 \caption{The pH value of colloidal samples in solution form is measured and compared to samples made under various conditions. Panel (a) shows pH values for samples prepared at different ablation powers for 20 minutes of ablation. Similarly, the variation in pH values for samples obtained at different ablation times at 1.5 W is compared in panel (b). Increasing ablation power and time result in lowering pH values, indicating that samples in solution are becoming more acidic.
}
\label{fgr:PS7}
\end{figure}

\newpage
\section{XRD data}

The XRD peaks are compared with the commercial power sample shown in Figure S8. The XRD patterns of the nanoparticles show a prominent peak at 14.39$^{\circ}$. This characteristic diffraction peak originates from the (002) plane of the $\mathrm{MoS_{2}}$, which is well aligned with the intense peak compared to its bulk form \cite{Zhou2014}. In addition to that, there were no peaks matching for $\mathrm{MoO_{3}}$, this suggests that XRD is an overall picture in which it is difficult to identify the minute changes.

\begin{figure}[h!]
\renewcommand{\figurename}{Figure S10}
 \renewcommand{\thefigure}{}
 \centering
\includegraphics[width=0.5\linewidth]{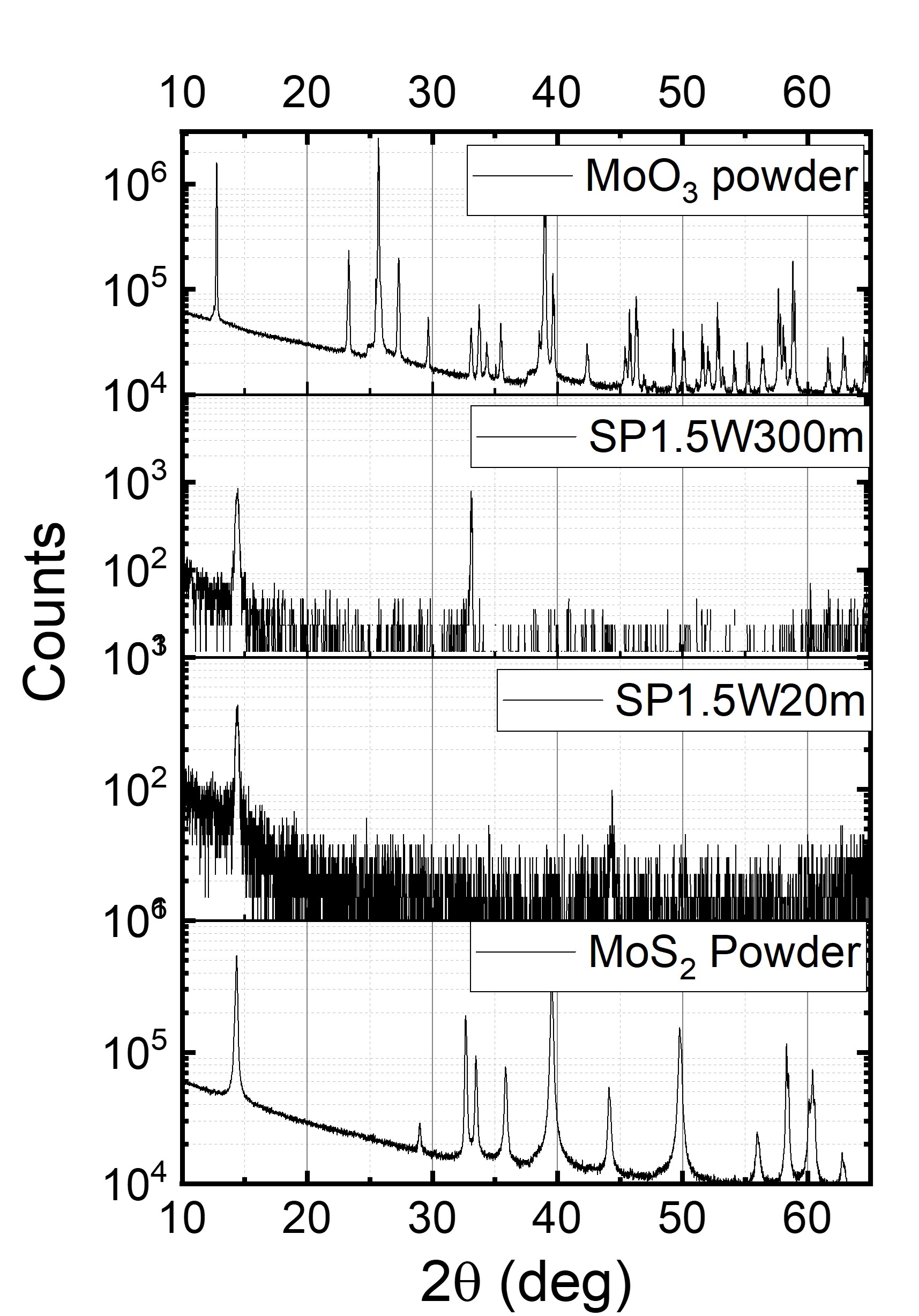}
 \caption{The semi-log plots of XRD data are compared for samples prepared for nanoparticles with commercial powders. }
\label{fgr:SP10}
\end{figure}

\end{suppinfo}

\begin{acknowledgement}
The financial support of the Dept. of Science and Technology (DST), Govt. of India, is gratefully acknowledged: A.S. was supported through the DST-INSPIRE scheme, P. K. N. through the financial Ramanujan Fellowship (SB/S2/RJN-043/2017), K. L. G. for the DST-INSPIRE faculty scheme (DST/INSPIRE/04/2016/001865), and S. K. for research grants through the Technology Development Board, Science and Engineering Research Board, and DST-DAAD bilateral scheme together with the German Academic Exchange Service (DAAD). M.S.R.R., S. K. and K. L. G.  acknowledge gratefully the support of the Ministry of Education, Govt. of India, through the Institute of Eminence (IoE) scheme, which funds the Quantum Center of Excellence for Diamond and Emergent Materials (QuCenDiEM); P. K. N. and K. L. G. likewise for IoE grants through the 2D Materials Research and Innovation group. M.S.R.R. acknowledges the support through the DST grant of the “Nano Functional Materials Technology Centre” (Grant: SRNM/NAT/02-2005). S. K. also acknowledges the partial support of the  Indo-French Center for Promotion of Academic Research (CEFIPRA).

\end{acknowledgement}

 \bibliography{References}

\providecommand{\latin}[1]{#1}
\makeatletter
\providecommand{\doi}
  {\begingroup\let\do\@makeother\dospecials
  \catcode`\{=1 \catcode`\}=2 \doi@aux}
\providecommand{\doi@aux}[1]{\endgroup\texttt{#1}}
\makeatother
\providecommand*\mcitethebibliography{\thebibliography}
\csname @ifundefined\endcsname{endmcitethebibliography}  {\let\endmcitethebibliography\endthebibliography}{}
\begin{mcitethebibliography}{68}
\providecommand*\natexlab[1]{#1}
\providecommand*\mciteSetBstSublistMode[1]{}
\providecommand*\mciteSetBstMaxWidthForm[2]{}
\providecommand*\mciteBstWouldAddEndPuncttrue
  {\def\EndOfBibitem{\unskip.}}
\providecommand*\mciteBstWouldAddEndPunctfalse
  {\let\EndOfBibitem\relax}
\providecommand*\mciteSetBstMidEndSepPunct[3]{}
\providecommand*\mciteSetBstSublistLabelBeginEnd[3]{}
\providecommand*\EndOfBibitem{}
\mciteSetBstSublistMode{f}
\mciteSetBstMaxWidthForm{subitem}{(\alph{mcitesubitemcount})}
\mciteSetBstSublistLabelBeginEnd
  {\mcitemaxwidthsubitemform\space}
  {\relax}
  {\relax}

\bibitem[Alivisatos(1996)]{Alivisatos933}
Alivisatos,~A.~P. Semiconductor Clusters, Nanocrystals, and Quantum Dots. \emph{Science} \textbf{1996}, \emph{271}, 933--937\relax
\mciteBstWouldAddEndPuncttrue
\mciteSetBstMidEndSepPunct{\mcitedefaultmidpunct}
{\mcitedefaultendpunct}{\mcitedefaultseppunct}\relax
\EndOfBibitem
\bibitem[Nirmal and Brus(1999)Nirmal, and Brus]{Nirmal1999}
Nirmal,~M.; Brus,~L. Luminescence Photophysics in Semiconductor Nanocrystals. \emph{Accounts of Chemical Research} \textbf{1999}, \emph{32}, 407--414\relax
\mciteBstWouldAddEndPuncttrue
\mciteSetBstMidEndSepPunct{\mcitedefaultmidpunct}
{\mcitedefaultendpunct}{\mcitedefaultseppunct}\relax
\EndOfBibitem
\bibitem[Yoffe(2002)]{Yoffe2002}
Yoffe,~A.~D. Low-dimensional systems: Quantum size effects and electronic properties of semiconductor microcrystallites (zero-dimensional systems) and some quasi-two-dimensional systems. \emph{Advances in Physics} \textbf{2002}, \emph{51}, 799--890\relax
\mciteBstWouldAddEndPuncttrue
\mciteSetBstMidEndSepPunct{\mcitedefaultmidpunct}
{\mcitedefaultendpunct}{\mcitedefaultseppunct}\relax
\EndOfBibitem
\bibitem[Arakawa(2002)]{Arakawa2002}
Arakawa,~Y. Progress in GaN-based quantum dots for optoelectronics applications. \emph{IEEE Journal of Selected Topics in Quantum Electronics} \textbf{2002}, \emph{8}, 823--832\relax
\mciteBstWouldAddEndPuncttrue
\mciteSetBstMidEndSepPunct{\mcitedefaultmidpunct}
{\mcitedefaultendpunct}{\mcitedefaultseppunct}\relax
\EndOfBibitem
\bibitem[Jin \latin{et~al.}(2015)Jin, Owour, Lei, and Ge]{JIN2015}
Jin,~Z.; Owour,~P.; Lei,~S.; Ge,~L. Graphene, graphene quantum dots and their applications in optoelectronics. \emph{Current Opinion in Colloid $\&$ Interface Science} \textbf{2015}, \emph{20}, 439--453\relax
\mciteBstWouldAddEndPuncttrue
\mciteSetBstMidEndSepPunct{\mcitedefaultmidpunct}
{\mcitedefaultendpunct}{\mcitedefaultseppunct}\relax
\EndOfBibitem
\bibitem[Hildebrandt \latin{et~al.}(2017)Hildebrandt, Spillmann, Algar, Pons, Stewart, Oh, Susumu, Díaz, Delehanty, and Medintz]{Hildebrandt2017}
Hildebrandt,~N.; Spillmann,~C.~M.; Algar,~W.~R.; Pons,~T.; Stewart,~M.~H.; Oh,~E.; Susumu,~K.; Díaz,~S.~A.; Delehanty,~J.~B.; Medintz,~I.~L. Energy Transfer with Semiconductor Quantum Dot Bioconjugates: A Versatile Platform for Biosensing, Energy Harvesting, and Other Developing Applications. \emph{Chemical Reviews} \textbf{2017}, \emph{117}, 536--711\relax
\mciteBstWouldAddEndPuncttrue
\mciteSetBstMidEndSepPunct{\mcitedefaultmidpunct}
{\mcitedefaultendpunct}{\mcitedefaultseppunct}\relax
\EndOfBibitem
\bibitem[Manikandan \latin{et~al.}(2019)Manikandan, Chen, Shen, Sher, Kuo, and Chueh]{MANIKANDAN2019}
Manikandan,~A.; Chen,~Y.-Z.; Shen,~C.-C.; Sher,~C.-W.; Kuo,~H.-C.; Chueh,~Y.-L. A critical review on two-dimensional quantum dots (2D QDs): From synthesis toward applications in energy and optoelectronics. \emph{Progress in Quantum Electronics} \textbf{2019}, \emph{68}, 100226\relax
\mciteBstWouldAddEndPuncttrue
\mciteSetBstMidEndSepPunct{\mcitedefaultmidpunct}
{\mcitedefaultendpunct}{\mcitedefaultseppunct}\relax
\EndOfBibitem
\bibitem[Yue \latin{et~al.}(2019)Yue, Li, McGuire, Hurley, and Wong]{Yue2019}
Yue,~S.; Li,~L.; McGuire,~S.~C.; Hurley,~N.; Wong,~S.~S. Metal chalcogenide quantum dot-sensitized 1D-based semiconducting heterostructures for optical-related applications. \emph{Energy Environ. Sci.} \textbf{2019}, \emph{12}, 1454--1494\relax
\mciteBstWouldAddEndPuncttrue
\mciteSetBstMidEndSepPunct{\mcitedefaultmidpunct}
{\mcitedefaultendpunct}{\mcitedefaultseppunct}\relax
\EndOfBibitem
\bibitem[Cotta(2020)]{Cotta2020}
Cotta,~M.~A. Quantum Dots and Their Applications: What Lies Ahead? \emph{ACS Applied Nano Materials} \textbf{2020}, \emph{3}, 4920--4924\relax
\mciteBstWouldAddEndPuncttrue
\mciteSetBstMidEndSepPunct{\mcitedefaultmidpunct}
{\mcitedefaultendpunct}{\mcitedefaultseppunct}\relax
\EndOfBibitem
\bibitem[Ghorai \latin{et~al.}(2017)Ghorai, Bayan, Gogurla, Midya, and Ray]{Arup2017}
Ghorai,~A.; Bayan,~S.; Gogurla,~N.; Midya,~A.; Ray,~S.~K. Highly Luminescent $\mathrm{WS_{2}}$ Quantum Dots/ZnO Heterojunctions for Light Emitting Devices. \emph{ACS Applied Materials \& Interfaces} \textbf{2017}, \emph{9}, 558--565\relax
\mciteBstWouldAddEndPuncttrue
\mciteSetBstMidEndSepPunct{\mcitedefaultmidpunct}
{\mcitedefaultendpunct}{\mcitedefaultseppunct}\relax
\EndOfBibitem
\bibitem[Bankar \latin{et~al.}(2017)Bankar, Khandare, Late, and More]{Bankar2017}
Bankar,~P.~K.; Khandare,~L.~N.; Late,~D.~J.; More,~M.~A. Enhanced Field Emission Performance of $\mathrm{MoO_{3}}$ Nanorods and $\mathrm{MoO_{3}}$-rGO Nanocomposite. \emph{ChemistrySelect} \textbf{2017}, \emph{2}, 10912--10917\relax
\mciteBstWouldAddEndPuncttrue
\mciteSetBstMidEndSepPunct{\mcitedefaultmidpunct}
{\mcitedefaultendpunct}{\mcitedefaultseppunct}\relax
\EndOfBibitem
\bibitem[Mukherjee \latin{et~al.}(2016)Mukherjee, Maiti, Katiyar, Das, and Ray]{Mukherjee2016}
Mukherjee,~S.; Maiti,~R.; Katiyar,~A.~K.; Das,~S.; Ray,~S.~K. Novel Colloidal $\mathrm{MoS_{2}}$ Quantum Dot Heterojunctions on Silicon Platforms for Multifunctional Optoelectronic Devices. \emph{Scientific Reports} \textbf{2016}, \emph{6}, 29016\relax
\mciteBstWouldAddEndPuncttrue
\mciteSetBstMidEndSepPunct{\mcitedefaultmidpunct}
{\mcitedefaultendpunct}{\mcitedefaultseppunct}\relax
\EndOfBibitem
\bibitem[Guo \latin{et~al.}(2020)Guo, Peng, Du, Shen, Li, Li, and Dong]{Guo2020}
Guo,~J.; Peng,~R.; Du,~H.; Shen,~Y.; Li,~Y.; Li,~J.; Dong,~G. The Application of Nano-$\mathrm{MoS_{2}}$ Quantum Dots as Liquid Lubricant Additive for Tribological Behavior Improvement. \emph{Nanomaterials} \textbf{2020}, \emph{10}\relax
\mciteBstWouldAddEndPuncttrue
\mciteSetBstMidEndSepPunct{\mcitedefaultmidpunct}
{\mcitedefaultendpunct}{\mcitedefaultseppunct}\relax
\EndOfBibitem
\bibitem[Yin \latin{et~al.}(2014)Yin, Zhang, Cai, Chen, Wong, Tay, Chai, Wu, Zeng, Zheng, Yang, and Zhang]{Yin2014}
Yin,~Z.; Zhang,~X.; Cai,~Y.; Chen,~J.; Wong,~J.; Tay,~Y.; Chai,~J.; Wu,~J.; Zeng,~Z.; Zheng,~B.; Yang,~H.; Zhang,~H. Preparation of $\mathrm{MoS_{2}-MoO_{3}}$ hybrid nanomaterials for light-emitting diodes. \emph{Angew Chem Int Ed Engl.} \textbf{2014}, \emph{53}, 12560 –12565\relax
\mciteBstWouldAddEndPuncttrue
\mciteSetBstMidEndSepPunct{\mcitedefaultmidpunct}
{\mcitedefaultendpunct}{\mcitedefaultseppunct}\relax
\EndOfBibitem
\bibitem[Roduner(2006)]{Rounder2006}
Roduner,~E. Size matters: why nanomaterials are different. \emph{Chem. Soc. Rev.} \textbf{2006}, \emph{35}, 583--592\relax
\mciteBstWouldAddEndPuncttrue
\mciteSetBstMidEndSepPunct{\mcitedefaultmidpunct}
{\mcitedefaultendpunct}{\mcitedefaultseppunct}\relax
\EndOfBibitem
\bibitem[Splendiani \latin{et~al.}(2010)Splendiani, Sun, Zhang, Li, Kim, Chim, Galli, and Wang]{Splendiani2010}
Splendiani,~A.; Sun,~L.; Zhang,~Y.; Li,~T.; Kim,~J.; Chim,~C.-Y.; Galli,~G.; Wang,~F. Emerging Photoluminescence in Monolayer $\mathrm{MoS_{2}}$. \emph{Nano Letters} \textbf{2010}, \emph{10}, 1271--1275\relax
\mciteBstWouldAddEndPuncttrue
\mciteSetBstMidEndSepPunct{\mcitedefaultmidpunct}
{\mcitedefaultendpunct}{\mcitedefaultseppunct}\relax
\EndOfBibitem
\bibitem[Nguyen \latin{et~al.}(2019)Nguyen, Dong, Yan, Zhao, and Le]{NGUYEN2019}
Nguyen,~V.; Dong,~Q.; Yan,~L.; Zhao,~N.; Le,~P.~H. Facile synthesis of photoluminescent $\mathrm{MoS_{2}}$ and $\mathrm{WS_{2}}$ quantum dots with strong surface-state emission. \emph{Journal of Luminescence} \textbf{2019}, \emph{214}, 116554\relax
\mciteBstWouldAddEndPuncttrue
\mciteSetBstMidEndSepPunct{\mcitedefaultmidpunct}
{\mcitedefaultendpunct}{\mcitedefaultseppunct}\relax
\EndOfBibitem
\bibitem[Lambora and Bhardwaj(2023)Lambora, and Bhardwaj]{Asha_2023}
Lambora,~S.; Bhardwaj,~A. Role of dielectric medium on optical behaviour of blue emitting colloidal $\mathrm{MoS_{2}}$ quantum Dots. \emph{Journal of Luminescence} \textbf{2023}, \emph{255}, 119598\relax
\mciteBstWouldAddEndPuncttrue
\mciteSetBstMidEndSepPunct{\mcitedefaultmidpunct}
{\mcitedefaultendpunct}{\mcitedefaultseppunct}\relax
\EndOfBibitem
\bibitem[Gao \latin{et~al.}(2017)Gao, Hao, Zheng, and Chen]{GaoStokes_2017}
Gao,~Z.; Hao,~Y.; Zheng,~M.; Chen,~Y. A fluorescent dye with large Stokes shift and high stability: synthesis and application to live cell imaging. \emph{RSC Adv.} \textbf{2017}, \emph{7}, 7604--7609\relax
\mciteBstWouldAddEndPuncttrue
\mciteSetBstMidEndSepPunct{\mcitedefaultmidpunct}
{\mcitedefaultendpunct}{\mcitedefaultseppunct}\relax
\EndOfBibitem
\bibitem[Jaiswal \latin{et~al.}(2022)Jaiswal, Girish, Behera, and De]{Jaiswal_Dual_2022}
Jaiswal,~K.; Girish,~Y.~R.; Behera,~P.; De,~M. Dual Role of $\mathrm{MoS_{2}}$ Quantum Dots in a Cross-Dehydrogenative Coupling Reaction. \emph{ACS Organic \& Inorganic Au} \textbf{2022}, \emph{2}, 205--213\relax
\mciteBstWouldAddEndPuncttrue
\mciteSetBstMidEndSepPunct{\mcitedefaultmidpunct}
{\mcitedefaultendpunct}{\mcitedefaultseppunct}\relax
\EndOfBibitem
\bibitem[Lin \latin{et~al.}(2015)Lin, Wang, Wu, Xu, Huang, and Zhang]{Lin2015}
Lin,~H.; Wang,~C.; Wu,~J.; Xu,~Z.; Huang,~Y.; Zhang,~C. Colloidal synthesis of $\mathrm{MoS_{2}}$ quantum dots: size-dependent tunable photoluminescence and bioimaging. \emph{New J. Chem.} \textbf{2015}, \emph{39}, 8492--8497\relax
\mciteBstWouldAddEndPuncttrue
\mciteSetBstMidEndSepPunct{\mcitedefaultmidpunct}
{\mcitedefaultendpunct}{\mcitedefaultseppunct}\relax
\EndOfBibitem
\bibitem[Portone \latin{et~al.}(2018)Portone, Romano, Fasano, Di~Corato, Camposeo, Fabbri, Cardarelli, Pisignano, and Persano]{Portone2018}
Portone,~A.; Romano,~L.; Fasano,~V.; Di~Corato,~R.; Camposeo,~A.; Fabbri,~F.; Cardarelli,~F.; Pisignano,~D.; Persano,~L. Low-defectiveness exfoliation of $\mathrm{MoS_{2}}$ nanoparticles and their embedment in hybrid light-emitting polymer nanofibers. \emph{Nanoscale} \textbf{2018}, \emph{10}, 21748--21754\relax
\mciteBstWouldAddEndPuncttrue
\mciteSetBstMidEndSepPunct{\mcitedefaultmidpunct}
{\mcitedefaultendpunct}{\mcitedefaultseppunct}\relax
\EndOfBibitem
\bibitem[Du \latin{et~al.}(2015)Du, Liu, Li, Al~Otaibi, Lv, and Zhang]{Du_RSC2015}
Du,~H.; Liu,~D.; Li,~M.; Al~Otaibi,~R.~L.; Lv,~R.; Zhang,~Y. Solvothermal synthesis of $\mathrm{MoS_{2}}$ nanospheres in DMF–water mixed solvents and their catalytic activity in hydrocracking of diphenylmethane. \emph{RSC Adv.} \textbf{2015}, \emph{5}, 79724--79728\relax
\mciteBstWouldAddEndPuncttrue
\mciteSetBstMidEndSepPunct{\mcitedefaultmidpunct}
{\mcitedefaultendpunct}{\mcitedefaultseppunct}\relax
\EndOfBibitem
\bibitem[Qiao \latin{et~al.}(2015)Qiao, Yan, Song, Zhang, He, Zhong, and Du]{Wen_ASS_2015}
Qiao,~W.; Yan,~S.; Song,~X.; Zhang,~X.; He,~X.; Zhong,~W.; Du,~Y. Luminescent monolayer $\mathrm{MoS_{2}}$ quantum dots produced by multi-exfoliation based on lithium intercalation. \emph{Applied Surface Science} \textbf{2015}, \emph{359}, 130 -- 136\relax
\mciteBstWouldAddEndPuncttrue
\mciteSetBstMidEndSepPunct{\mcitedefaultmidpunct}
{\mcitedefaultendpunct}{\mcitedefaultseppunct}\relax
\EndOfBibitem
\bibitem[Ali \latin{et~al.}(2022)Ali, Subhan, Ayaz, Hassan, Byeon, Kim, and Bungau]{Ali2022}
Ali,~L.; Subhan,~F.; Ayaz,~M.; Hassan,~S.; Byeon,~C.; Kim,~J.; Bungau,~S. Exfoliation of $\mathrm{MoS_{2}}$ Quantum Dots: Recent Progress and Challenges. \emph{Nanomaterials} \textbf{2022}, \emph{12}\relax
\mciteBstWouldAddEndPuncttrue
\mciteSetBstMidEndSepPunct{\mcitedefaultmidpunct}
{\mcitedefaultendpunct}{\mcitedefaultseppunct}\relax
\EndOfBibitem
\bibitem[Lambora and Bhardwaj(2023)Lambora, and Bhardwaj]{LAMBORA2023}
Lambora,~S.; Bhardwaj,~A. Role of dielectric medium on optical behaviour of blue emitting colloidal $\mathrm{MoS_{2}}$ quantum Dots. \emph{Journal of Luminescence} \textbf{2023}, \emph{255}, 119598\relax
\mciteBstWouldAddEndPuncttrue
\mciteSetBstMidEndSepPunct{\mcitedefaultmidpunct}
{\mcitedefaultendpunct}{\mcitedefaultseppunct}\relax
\EndOfBibitem
\bibitem[Gan \latin{et~al.}(2015)Gan, Liu, Wu, Hao, Shan, Wu, and Chu]{Gan_2015}
Gan,~Z.~X.; Liu,~L.~Z.; Wu,~H.~Y.; Hao,~Y.~L.; Shan,~Y.; Wu,~X.~L.; Chu,~P.~K. Quantum confinement effects across two-dimensional planes in $\mathrm{MoS_{2}}$ quantum dots. \emph{Applied Physics Letters} \textbf{2015}, \emph{106}, 233113\relax
\mciteBstWouldAddEndPuncttrue
\mciteSetBstMidEndSepPunct{\mcitedefaultmidpunct}
{\mcitedefaultendpunct}{\mcitedefaultseppunct}\relax
\EndOfBibitem
\bibitem[Li \latin{et~al.}(2017)Li, Jiang, Li, Ran, Zuo, Wang, Qu, Zhao, Cheng, and Lu]{Li_SR_2017}
Li,~B.; Jiang,~L.; Li,~X.; Ran,~P.; Zuo,~P.; Wang,~A.; Qu,~L.; Zhao,~Y.; Cheng,~Z.; Lu,~Y. Preparation of Monolayer $\mathrm{MoS_{2}}$ Quantum Dots using Temporally Shaped Femtosecond Laser Ablation of Bulk $\mathrm{MoS_{2}}$ Targets in Water. \emph{Scientific Reports} \textbf{2017}, \emph{7}, 11182\relax
\mciteBstWouldAddEndPuncttrue
\mciteSetBstMidEndSepPunct{\mcitedefaultmidpunct}
{\mcitedefaultendpunct}{\mcitedefaultseppunct}\relax
\EndOfBibitem
\bibitem[Yin \latin{et~al.}(2019)Yin, Bai, Zhang, Zhang, Gao, and Yu]{Yin2019}
Yin,~W.; Bai,~X.; Zhang,~X.; Zhang,~J.; Gao,~X.; Yu,~W.~W. Multicolor Light-Emitting Diodes with $\mathrm{MoS_{2}}$ Quantum Dots. \emph{Particle \& Particle Systems Characterization} \textbf{2019}, \emph{36}, 1800362\relax
\mciteBstWouldAddEndPuncttrue
\mciteSetBstMidEndSepPunct{\mcitedefaultmidpunct}
{\mcitedefaultendpunct}{\mcitedefaultseppunct}\relax
\EndOfBibitem
\bibitem[Xu \latin{et~al.}(2019)Xu, Yan, Li, and Xu]{Xu_SR_2019}
Xu,~Y.; Yan,~L.; Li,~X.; Xu,~H. Fabrication of transition metal dichalcogenides quantum dots based on femtosecond laser ablation. \emph{Scientific Reports} \textbf{2019}, \emph{9}, 2931\relax
\mciteBstWouldAddEndPuncttrue
\mciteSetBstMidEndSepPunct{\mcitedefaultmidpunct}
{\mcitedefaultendpunct}{\mcitedefaultseppunct}\relax
\EndOfBibitem
\bibitem[Sunitha \latin{et~al.}(2018)Sunitha, Hajara, Shaji, Jayaraj, and Saji]{SUNITHA2018}
Sunitha,~A.; Hajara,~P.; Shaji,~M.; Jayaraj,~M.; Saji,~K. Luminescent $\mathrm{MoS_{2}}$ quantum dots with reverse saturable absorption prepared by pulsed laser ablation. \emph{Journal of Luminescence} \textbf{2018}, \emph{203}, 313 -- 321\relax
\mciteBstWouldAddEndPuncttrue
\mciteSetBstMidEndSepPunct{\mcitedefaultmidpunct}
{\mcitedefaultendpunct}{\mcitedefaultseppunct}\relax
\EndOfBibitem
\bibitem[Alsaif \latin{et~al.}(2014)Alsaif, Latham, Field, Yao, Medehkar, Beane, Kaner, Russo, Ou, and Kalantar-zadeh]{Alsaif2014}
Alsaif,~M. M. Y.~A.; Latham,~K.; Field,~M.~R.; Yao,~D.~D.; Medehkar,~N.~V.; Beane,~G.~A.; Kaner,~R.~B.; Russo,~S.~P.; Ou,~J.~Z.; Kalantar-zadeh,~K. Tunable Plasmon Resonances in Two-Dimensional Molybdenum Oxide Nanoflakes. \emph{Advanced Materials} \textbf{2014}, \emph{26}, 3931--3937\relax
\mciteBstWouldAddEndPuncttrue
\mciteSetBstMidEndSepPunct{\mcitedefaultmidpunct}
{\mcitedefaultendpunct}{\mcitedefaultseppunct}\relax
\EndOfBibitem
\bibitem[Annu{\v{s}}ov{\'{a}} \latin{et~al.}(2020)Annu{\v{s}}ov{\'{a}}, Bod{\'{\i}}k, Hagara, Kotl{\'{a}}r, Halahovets, Mi{\v{c}}u{\v{s}}{\'{\i}}k, Chlp{\'{\i}}k, Cir{\'{a}}k, Hofbauerov{\'{a}}, Jergel, Majkov{\'{a}}, and {\v{S}}iffalovi{\v{c}}]{Annu2020}
Annu{\v{s}}ov{\'{a}},~A.; Bod{\'{\i}}k,~M.; Hagara,~J.; Kotl{\'{a}}r,~M.; Halahovets,~Y.; Mi{\v{c}}u{\v{s}}{\'{\i}}k,~M.; Chlp{\'{\i}}k,~J.; Cir{\'{a}}k,~J.; Hofbauerov{\'{a}},~M.; Jergel,~M.; Majkov{\'{a}},~E.; {\v{S}}iffalovi{\v{c}},~P. On the extraction of $\mathrm{MoO_{x}}$ photothermally active nanoparticles by gel filtration from byproduct of few-layer $\mathrm{MoS_{2}}$ exfoliation. \emph{Nanotechnology} \textbf{2020}, \emph{32}, 045708\relax
\mciteBstWouldAddEndPuncttrue
\mciteSetBstMidEndSepPunct{\mcitedefaultmidpunct}
{\mcitedefaultendpunct}{\mcitedefaultseppunct}\relax
\EndOfBibitem
\bibitem[Zamora-Romero \latin{et~al.}(2020)Zamora-Romero, Camacho-Lopez, Vilchis-Nestor, Castrejon-Sanchez, Aguilar, Camacho-Lopez, and Camacho-Lopez]{ZAMORAROMERO2020}
Zamora-Romero,~N.; Camacho-Lopez,~M.~A.; Vilchis-Nestor,~A.~R.; Castrejon-Sanchez,~V.~H.; Aguilar,~G.; Camacho-Lopez,~S.; Camacho-Lopez,~M. Synthesis of molybdenum oxide nanoparticles by nanosecond laser ablation. \emph{Materials Chemistry and Physics} \textbf{2020}, \emph{240}, 122163\relax
\mciteBstWouldAddEndPuncttrue
\mciteSetBstMidEndSepPunct{\mcitedefaultmidpunct}
{\mcitedefaultendpunct}{\mcitedefaultseppunct}\relax
\EndOfBibitem
\bibitem[Li \latin{et~al.}(2021)Li, Xu, Huang, Lou, and Li]{Li_2021}
Li,~J.; Xu,~X.; Huang,~B.; Lou,~Z.; Li,~B. Light-Induced In Situ Formation of a Nonmetallic Plasmonic $\mathrm{MoS_{2}}$/$\mathrm{MoO_{3-x}}$ Heterostructure with Efficient Charge Transfer for CO2 Reduction and SERS Detection. \emph{ACS Applied Materials \& Interfaces} \textbf{2021}, \emph{13}, 10047--10053\relax
\mciteBstWouldAddEndPuncttrue
\mciteSetBstMidEndSepPunct{\mcitedefaultmidpunct}
{\mcitedefaultendpunct}{\mcitedefaultseppunct}\relax
\EndOfBibitem
\bibitem[Eda \latin{et~al.}(2011)Eda, Yamaguchi, Voiry, Fujita, Chen, and Chhowalla]{Eda2011}
Eda,~G.; Yamaguchi,~H.; Voiry,~D.; Fujita,~T.; Chen,~M.; Chhowalla,~M. Photoluminescence from Chemically Exfoliated $\mathrm{MoS_{2}}$. \emph{Nano Letters} \textbf{2011}, \emph{11}, 5111--5116\relax
\mciteBstWouldAddEndPuncttrue
\mciteSetBstMidEndSepPunct{\mcitedefaultmidpunct}
{\mcitedefaultendpunct}{\mcitedefaultseppunct}\relax
\EndOfBibitem
\bibitem[Doolen \latin{et~al.}(1998)Doolen, Laitinen, Parsapour, and Kelley]{Doolen1998}
Doolen,~R.; Laitinen,~R.; Parsapour,~F.; Kelley,~D.~F. Trap State Dynamics in $\mathrm{MoS_{2}}$ Nanoclusters. \emph{The Journal of Physical Chemistry B} \textbf{1998}, \emph{102}, 3906--3911\relax
\mciteBstWouldAddEndPuncttrue
\mciteSetBstMidEndSepPunct{\mcitedefaultmidpunct}
{\mcitedefaultendpunct}{\mcitedefaultseppunct}\relax
\EndOfBibitem
\bibitem[Gopalakrishnan \latin{et~al.}(2015)Gopalakrishnan, Damien, Li, Gullappalli, Pillai, Ajayan, and Shaijumon]{Gopalakrishnan2015}
Gopalakrishnan,~D.; Damien,~D.; Li,~B.; Gullappalli,~H.; Pillai,~V.~K.; Ajayan,~P.~M.; Shaijumon,~M.~M. Electrochemical synthesis of luminescent $\mathrm{MoS_{2}}$ quantum dots. \emph{Chem. Commun.} \textbf{2015}, \emph{51}, 6293--6296\relax
\mciteBstWouldAddEndPuncttrue
\mciteSetBstMidEndSepPunct{\mcitedefaultmidpunct}
{\mcitedefaultendpunct}{\mcitedefaultseppunct}\relax
\EndOfBibitem
\bibitem[An \latin{et~al.}(2020)An, Park, Lee, Bang, Nguyen, Kim, Kim, Jeong, and Jeong]{Sung2020}
An,~S.-J.; Park,~D.~Y.; Lee,~C.; Bang,~S.; Nguyen,~D.~A.; Kim,~S.~H.; Kim,~H.~Y.; Jeong,~H.~J.; Jeong,~M.~S. Facile preparation of molybdenum disulfide quantum dots using a femtosecond laser. \emph{Applied Surface Science} \textbf{2020}, \emph{511}, 145507\relax
\mciteBstWouldAddEndPuncttrue
\mciteSetBstMidEndSepPunct{\mcitedefaultmidpunct}
{\mcitedefaultendpunct}{\mcitedefaultseppunct}\relax
\EndOfBibitem
\bibitem[Li \latin{et~al.}(2012)Li, Zhang, Yap, Tay, Edwin, Olivier, and Baillargeat]{Hong2012}
Li,~H.; Zhang,~Q.; Yap,~C. C.~R.; Tay,~B.~K.; Edwin,~T. H.~T.; Olivier,~A.; Baillargeat,~D. From Bulk to Monolayer $\mathrm{MoS_{2}}$: Evolution of Raman Scattering. \emph{Advanced Functional Materials} \textbf{2012}, \emph{22}, 1385--1390\relax
\mciteBstWouldAddEndPuncttrue
\mciteSetBstMidEndSepPunct{\mcitedefaultmidpunct}
{\mcitedefaultendpunct}{\mcitedefaultseppunct}\relax
\EndOfBibitem
\bibitem[Gnanasekar \latin{et~al.}(2018)Gnanasekar, Periyanagounder, Nallathambi, Subramani, Palanisamy, and Kulandaivel]{Gnanasekar2018}
Gnanasekar,~P.; Periyanagounder,~D.; Nallathambi,~A.; Subramani,~S.; Palanisamy,~M.; Kulandaivel,~J. Promoter-free synthesis of monolayer $\mathrm{MoS_{2}}$ by chemical vapour deposition. \emph{CrystEngComm} \textbf{2018}, \emph{20}, 4249--4257\relax
\mciteBstWouldAddEndPuncttrue
\mciteSetBstMidEndSepPunct{\mcitedefaultmidpunct}
{\mcitedefaultendpunct}{\mcitedefaultseppunct}\relax
\EndOfBibitem
\bibitem[de~Barros~Santos \latin{et~al.}(2012)de~Barros~Santos, Sigoli, and Mazali]{SANTOS2012}
de~Barros~Santos,~E.; Sigoli,~F.~A.; Mazali,~I.~O. Structural evolution in crystalline $\mathrm{MoO_{3}}$ nanoparticles with tunable size. \emph{Journal of Solid State Chemistry} \textbf{2012}, \emph{190}, 80--84\relax
\mciteBstWouldAddEndPuncttrue
\mciteSetBstMidEndSepPunct{\mcitedefaultmidpunct}
{\mcitedefaultendpunct}{\mcitedefaultseppunct}\relax
\EndOfBibitem
\bibitem[Lee \latin{et~al.}(2019)Lee, Park, Kim, Kim, and Kim]{Lee2019}
Lee,~J.~S.; Park,~C.-S.; Kim,~T.~Y.; Kim,~Y.~S.; Kim,~E.~K. Characteristics of p-Type Conduction in P-Doped $\mathrm{MoS_{2}}$ by Phosphorous Pentoxide during Chemical Vapor Deposition. \emph{Nanomaterials} \textbf{2019}, \emph{9}\relax
\mciteBstWouldAddEndPuncttrue
\mciteSetBstMidEndSepPunct{\mcitedefaultmidpunct}
{\mcitedefaultendpunct}{\mcitedefaultseppunct}\relax
\EndOfBibitem
\bibitem[Ko \latin{et~al.}(2016)Ko, Jeong, Kim, Lee, Kim, Lee, Ryu, Park, Kim, Lee, Lee, Lee, and Ryu]{ko_2017}
Ko,~T.~Y.; Jeong,~A.; Kim,~W.; Lee,~J.; Kim,~Y.; Lee,~J.~E.; Ryu,~G.~H.; Park,~K.; Kim,~D.; Lee,~Z.; Lee,~M.~H.; Lee,~C.; Ryu,~S. On-stack two-dimensional conversion of $\mathrm{MoS_{2}}$ into $\mathrm{MoO_{3}}$. \emph{2D Materials} \textbf{2016}, \emph{4}, 014003\relax
\mciteBstWouldAddEndPuncttrue
\mciteSetBstMidEndSepPunct{\mcitedefaultmidpunct}
{\mcitedefaultendpunct}{\mcitedefaultseppunct}\relax
\EndOfBibitem
\bibitem[Chakraborty \latin{et~al.}(2012)Chakraborty, Bera, Muthu, Bhowmick, Waghmare, and Sood]{Chakraborty_PRL2012}
Chakraborty,~B.; Bera,~A.; Muthu,~D. V.~S.; Bhowmick,~S.; Waghmare,~U.~V.; Sood,~A.~K. Symmetry-dependent phonon renormalization in monolayer MoS${}_{2}$ transistor. \emph{Phys. Rev. B} \textbf{2012}, \emph{85}, 161403\relax
\mciteBstWouldAddEndPuncttrue
\mciteSetBstMidEndSepPunct{\mcitedefaultmidpunct}
{\mcitedefaultendpunct}{\mcitedefaultseppunct}\relax
\EndOfBibitem
\bibitem[Sharma \latin{et~al.}(2020)Sharma, Rao, Singh, and Vasa]{G_Sharma2020}
Sharma,~G.; Rao,~S.~M.; Singh,~B.~P.; Vasa,~P. Optically tunable charge carrier injection in monolayer $\mathrm{MoS_{2}}$. \emph{Applied Physics A} \textbf{2020}, \emph{126}, 663\relax
\mciteBstWouldAddEndPuncttrue
\mciteSetBstMidEndSepPunct{\mcitedefaultmidpunct}
{\mcitedefaultendpunct}{\mcitedefaultseppunct}\relax
\EndOfBibitem
\bibitem[Kumar \latin{et~al.}(2020)Kumar, Singh, and Reddy]{Prabhat_2020}
Kumar,~P.; Singh,~M.; Reddy,~G.~B. Oxidized Core–Shell $\mathrm{MoO_{2}}$–$\mathrm{MoS_{2}}$ Nanostructured Thin Films for Hydrogen Evolution. \emph{ACS Applied Nano Materials} \textbf{2020}, \emph{3}, 711--723\relax
\mciteBstWouldAddEndPuncttrue
\mciteSetBstMidEndSepPunct{\mcitedefaultmidpunct}
{\mcitedefaultendpunct}{\mcitedefaultseppunct}\relax
\EndOfBibitem
\bibitem[Afanasiev and Lorentz(2019)Afanasiev, and Lorentz]{Afanasiev2019}
Afanasiev,~P.; Lorentz,~C. Oxidation of Nanodispersed $\mathrm{MoS_{2}}$ in Ambient Air: The Products and the Mechanistic Steps. \emph{The Journal of Physical Chemistry C} \textbf{2019}, \emph{123}, 7486--7494\relax
\mciteBstWouldAddEndPuncttrue
\mciteSetBstMidEndSepPunct{\mcitedefaultmidpunct}
{\mcitedefaultendpunct}{\mcitedefaultseppunct}\relax
\EndOfBibitem
\bibitem[Chikan and Kelley(2002)Chikan, and Kelley]{Chikan2002}
Chikan,~V.; Kelley,~D.~F. Size-Dependent Spectroscopy of $\mathrm{MoS_{2}}$ Nanoclusters. \emph{The Journal of Physical Chemistry B} \textbf{2002}, \emph{106}, 3794--3804\relax
\mciteBstWouldAddEndPuncttrue
\mciteSetBstMidEndSepPunct{\mcitedefaultmidpunct}
{\mcitedefaultendpunct}{\mcitedefaultseppunct}\relax
\EndOfBibitem
\bibitem[Hariharan and Karthikeyan(2016)Hariharan, and Karthikeyan]{Hariharan2016}
Hariharan,~S.; Karthikeyan,~B. Optical and surface band bending mediated fluorescence sensing properties of $\mathrm{MoS_{2}}$ quantum dots. \emph{RSC Adv.} \textbf{2016}, \emph{6}, 101770--101777\relax
\mciteBstWouldAddEndPuncttrue
\mciteSetBstMidEndSepPunct{\mcitedefaultmidpunct}
{\mcitedefaultendpunct}{\mcitedefaultseppunct}\relax
\EndOfBibitem
\bibitem[Ha \latin{et~al.}(2014)Ha, Han, Choi, Park, and Seo]{Ha2014}
Ha,~H.~D.; Han,~D.~J.; Choi,~J.~S.; Park,~M.; Seo,~T.~S. Photoluminescence: Dual Role of Blue Luminescent $\mathrm{MoS_{2}}$ Quantum Dots in Fluorescence Resonance Energy Transfer Phenomenon. \emph{Small} \textbf{2014}, \emph{10}, 3814--3814\relax
\mciteBstWouldAddEndPuncttrue
\mciteSetBstMidEndSepPunct{\mcitedefaultmidpunct}
{\mcitedefaultendpunct}{\mcitedefaultseppunct}\relax
\EndOfBibitem
\bibitem[Wang and Ni(2014)Wang, and Ni]{Wang_AnlChem_2014}
Wang,~Y.; Ni,~Y. Molybdenum Disulfide Quantum Dots as a Photoluminescence Sensing Platform for 2,4,6-Trinitrophenol Detection. \emph{Analytical Chemistry} \textbf{2014}, \emph{86}, 7463--7470\relax
\mciteBstWouldAddEndPuncttrue
\mciteSetBstMidEndSepPunct{\mcitedefaultmidpunct}
{\mcitedefaultendpunct}{\mcitedefaultseppunct}\relax
\EndOfBibitem
\bibitem[Peng \latin{et~al.}(2016)Peng, Yu, Liu, Liu, Liang, Bi, Deng, Sum, and Loh]{Peng_2016}
Peng,~B.; Yu,~G.; Liu,~X.; Liu,~B.; Liang,~X.; Bi,~L.; Deng,~L.; Sum,~T.~C.; Loh,~K.~P. Ultrafast charge transfer in $\mathrm{{MoS}_{2} /WSe_{2}}$ p-n Heterojunction. \emph{2D Materials} \textbf{2016}, \emph{3}, 025020\relax
\mciteBstWouldAddEndPuncttrue
\mciteSetBstMidEndSepPunct{\mcitedefaultmidpunct}
{\mcitedefaultendpunct}{\mcitedefaultseppunct}\relax
\EndOfBibitem
\bibitem[Santiago \latin{et~al.}(2020)Santiago, Wang, Chen, Hsu, Wu, Hsu, Cheng, Lin, Feria, Chou, and Shen]{Santiago2020}
Santiago,~S. R. M.~S.; Wang,~H.-J.; Chen,~Y.-T.; Hsu,~I.-J.; Wu,~C.-B.; Hsu,~K.-M.; Cheng,~M.-C.; Lin,~T.-N.; Feria,~D.~N.; Chou,~W.-C.; Shen,~J.-L. Density-Dependent Carrier Recombination in $\mathrm{MoS_{2}}$ Quantum Dots and Its Implications for Luminescence Sensing of Ammonium Hydroxide. \emph{ACS Applied Nano Materials} \textbf{2020}, \emph{3}, 11630--11637\relax
\mciteBstWouldAddEndPuncttrue
\mciteSetBstMidEndSepPunct{\mcitedefaultmidpunct}
{\mcitedefaultendpunct}{\mcitedefaultseppunct}\relax
\EndOfBibitem
\bibitem[Liu \latin{et~al.}(2021)Liu, Li, Zhang, and Lu]{Liu_2021}
Liu,~J.; Li,~Z.; Zhang,~X.; Lu,~G. Unraveling energy and charge transfer in type-II van der Waals heterostructures. \emph{npj Computational Materials} \textbf{2021}, \emph{7}, 191\relax
\mciteBstWouldAddEndPuncttrue
\mciteSetBstMidEndSepPunct{\mcitedefaultmidpunct}
{\mcitedefaultendpunct}{\mcitedefaultseppunct}\relax
\EndOfBibitem
\bibitem[Zhao \latin{et~al.}(2017)Zhao, Amani, Lien, Ahn, Kiriya, Mastandrea, Ager, Yablonovitch, Chrzan, and Javey]{Zhao2017}
Zhao,~P.; Amani,~M.; Lien,~D.-H.; Ahn,~G.~H.; Kiriya,~D.; Mastandrea,~J.~P.; Ager,~J.~W.; Yablonovitch,~E.; Chrzan,~D.~C.; Javey,~A. Measuring the Edge Recombination Velocity of Monolayer Semiconductors. \emph{Nano Letters} \textbf{2017}, \emph{17}, 5356--5360\relax
\mciteBstWouldAddEndPuncttrue
\mciteSetBstMidEndSepPunct{\mcitedefaultmidpunct}
{\mcitedefaultendpunct}{\mcitedefaultseppunct}\relax
\EndOfBibitem
\bibitem[Wang \latin{et~al.}(2018)Wang, Zhang, Liu, Li, Liu, Luo, and Ge]{Wang2018}
Wang,~T.; Zhang,~Y.; Liu,~Y.; Li,~J.; Liu,~D.; Luo,~J.; Ge,~K. Layer-Number-Dependent Exciton Recombination Behaviors of $\mathrm{MoS_{2}}$ Determined by Fluorescence-Lifetime Imaging Microscopy. \emph{The Journal of Physical Chemistry C} \textbf{2018}, \emph{122}, 18651--18658\relax
\mciteBstWouldAddEndPuncttrue
\mciteSetBstMidEndSepPunct{\mcitedefaultmidpunct}
{\mcitedefaultendpunct}{\mcitedefaultseppunct}\relax
\EndOfBibitem
\bibitem[Bhattacharya \latin{et~al.}(2020)Bhattacharya, Mukherjee, Mitra, and Ray]{Bhattacharya_2020}
Bhattacharya,~D.; Mukherjee,~S.; Mitra,~R.~K.; Ray,~S.~K. Size-dependent optical properties of $\mathrm{{MoS}_2}$ nanoparticles and their photo-catalytic applications. \emph{Nanotechnology} \textbf{2020}, \emph{31}, 145701\relax
\mciteBstWouldAddEndPuncttrue
\mciteSetBstMidEndSepPunct{\mcitedefaultmidpunct}
{\mcitedefaultendpunct}{\mcitedefaultseppunct}\relax
\EndOfBibitem
\bibitem[Nguyen \latin{et~al.}(2015)Nguyen, Si, Yan, and Hou]{NGUYEN2015}
Nguyen,~V.; Si,~J.; Yan,~L.; Hou,~X. Electron–hole recombination dynamics in carbon nanodots. \emph{Carbon} \textbf{2015}, \emph{95}, 659--663\relax
\mciteBstWouldAddEndPuncttrue
\mciteSetBstMidEndSepPunct{\mcitedefaultmidpunct}
{\mcitedefaultendpunct}{\mcitedefaultseppunct}\relax
\EndOfBibitem
\bibitem[Tanoh \latin{et~al.}(2020)Tanoh, Gauriot, Delport, Xiao, Pandya, Sung, Allardice, Li, Williams, Baldwin, Stranks, and Rao]{Tanoh2020}
Tanoh,~A. O.~A.; Gauriot,~N.; Delport,~G.; Xiao,~J.; Pandya,~R.; Sung,~J.; Allardice,~J.; Li,~Z.; Williams,~C.~A.; Baldwin,~A.; Stranks,~S.~D.; Rao,~A. Directed Energy Transfer from Monolayer $\mathrm{WS_{2}}$ to Near-Infrared Emitting PbS–CdS Quantum Dots. \emph{ACS Nano} \textbf{2020}, \emph{14}, 15374--15384\relax
\mciteBstWouldAddEndPuncttrue
\mciteSetBstMidEndSepPunct{\mcitedefaultmidpunct}
{\mcitedefaultendpunct}{\mcitedefaultseppunct}\relax
\EndOfBibitem
\bibitem[Li \latin{et~al.}(2020)Li, Yao, Wu, Zhang, Xing, Niu, Yan, Yu, Liu, and Wang]{Li_Ptype2020}
Li,~M.; Yao,~J.; Wu,~X.; Zhang,~S.; Xing,~B.; Niu,~X.; Yan,~X.; Yu,~Y.; Liu,~Y.; Wang,~Y. P-type Doping in Large-Area Monolayer $\mathrm{MoS_{2}}$ by Chemical Vapor Deposition. \emph{ACS Applied Materials \& Interfaces} \textbf{2020}, \emph{12}, 6276--6282\relax
\mciteBstWouldAddEndPuncttrue
\mciteSetBstMidEndSepPunct{\mcitedefaultmidpunct}
{\mcitedefaultendpunct}{\mcitedefaultseppunct}\relax
\EndOfBibitem
\bibitem[Santosh \latin{et~al.}(2016)Santosh, Longo, Addou, Wallace, and Cho]{Santosh2016}
Santosh,~K.~C.; Longo,~R.~C.; Addou,~R.; Wallace,~R.~M.; Cho,~K. Electronic properties of $\mathrm{MoS_{2}/MoO_{x}}$ interfaces: Implications in Tunnel Field Effect Transistors and Hole Contacts. \emph{Scientific Reports} \textbf{2016}, \emph{6}, 33562\relax
\mciteBstWouldAddEndPuncttrue
\mciteSetBstMidEndSepPunct{\mcitedefaultmidpunct}
{\mcitedefaultendpunct}{\mcitedefaultseppunct}\relax
\EndOfBibitem
\bibitem[Hojamberdiev \latin{et~al.}(2019)Hojamberdiev, Zhu, Kumar, Wang, and Gao]{Mirabbos2019}
Hojamberdiev,~M.; Zhu,~G.; Kumar,~M.; Wang,~M.; Gao,~J. $\mathrm{MoS_{2}}$ quantum dots-modified porous $\mathrm{\beta -Bi_{2}O_{3}}$ microspheres with enhanced visible-light-induced photocatalytic activity for Bisphenol A degradation and NO removal. \emph{Journal of Materials Science: Materials in Electronics} \textbf{2019}, \emph{30}, 2610--2621\relax
\mciteBstWouldAddEndPuncttrue
\mciteSetBstMidEndSepPunct{\mcitedefaultmidpunct}
{\mcitedefaultendpunct}{\mcitedefaultseppunct}\relax
\EndOfBibitem
\bibitem[Suneel~Kumar and Krishnan(2020)Suneel~Kumar, and Krishnan]{Suneel2020}
Suneel~Kumar,~A.~K.,~Ajay~Kumar; Krishnan,~V. Nanoscale zinc oxide based heterojunctions as visible light active photocatalysts for hydrogen energy and environmental remediation. \emph{Catalysis Reviews} \textbf{2020}, \emph{62}, 346--405\relax
\mciteBstWouldAddEndPuncttrue
\mciteSetBstMidEndSepPunct{\mcitedefaultmidpunct}
{\mcitedefaultendpunct}{\mcitedefaultseppunct}\relax
\EndOfBibitem
\bibitem[Jagminas \latin{et~al.}(2016)Jagminas, Niaura, Žalnėravičius, Trusovas, Račiukaitis, and Jasulaitiene]{Jagminas2016}
Jagminas,~A.; Niaura,~G.; Žalnėravičius,~R.; Trusovas,~R.; Račiukaitis,~G.; Jasulaitiene,~V. Laser Light Induced Transformation of Molybdenum Disulphide-Based Nanoplatelet Arrays. \emph{Scientific Reports} \textbf{2016}, \emph{6}, 37514\relax
\mciteBstWouldAddEndPuncttrue
\mciteSetBstMidEndSepPunct{\mcitedefaultmidpunct}
{\mcitedefaultendpunct}{\mcitedefaultseppunct}\relax
\EndOfBibitem
\bibitem[{\v{s}}ov{\'{a}} \latin{et~al.}(2020){\v{s}}ov{\'{a}}, Bod{\'{\i}}k, Hagara, Kotl{\'{a}}r, Halahovets, Mi{\v{c}}u{\v{s}}{\'{\i}}k, Chlp{\'{\i}}k, Cir{\'{a}}k, Hofbauerov{\'{a}}, Jergel, Majkov{\'{a}}, and {\v{S}}iffalovi{\v{c}}]{Annu_ov__2020}
{\v{s}}ov{\'{a}},~A.~A.; Bod{\'{\i}}k,~M.; Hagara,~J.; Kotl{\'{a}}r,~M.; Halahovets,~Y.; Mi{\v{c}}u{\v{s}}{\'{\i}}k,~M.; Chlp{\'{\i}}k,~J.; Cir{\'{a}}k,~J.; Hofbauerov{\'{a}},~M.; Jergel,~M.; Majkov{\'{a}},~E.; {\v{S}}iffalovi{\v{c}},~P. On the extraction of $\mathrm{MoO_{x}}$ photothermally active nanoparticles by gel filtration from a byproduct of few-layer $MoS_{2}$ exfoliation. \emph{Nanotechnology} \textbf{2020}, \emph{32}, 045708\relax
\mciteBstWouldAddEndPuncttrue
\mciteSetBstMidEndSepPunct{\mcitedefaultmidpunct}
{\mcitedefaultendpunct}{\mcitedefaultseppunct}\relax
\EndOfBibitem
\bibitem[Zhou \latin{et~al.}(2014)Zhou, Xu, Yu, Feng, Zhang, Hu, and Zhou]{Zhou2014}
Zhou,~G.; Xu,~X.; Yu,~J.; Feng,~B.; Zhang,~Y.; Hu,~J.; Zhou,~Y. Vertically aligned $\mathrm{MoS_{2}}$/$\mathrm{MoO_{x}}$ heterojunction nanosheets for enhanced visible-light photocatalytic activity and photostability. \emph{CrystEngComm} \textbf{2014}, \emph{16}, 9025--9032\relax
\mciteBstWouldAddEndPuncttrue
\mciteSetBstMidEndSepPunct{\mcitedefaultmidpunct}
{\mcitedefaultendpunct}{\mcitedefaultseppunct}\relax
\EndOfBibitem
\end{mcitethebibliography}

\end{document}